# Correlative 3D Mapping of Structure, Composition, and Valence State Dynamics in Battery Cathodes via Simultaneous ADF-EDS-EELS Tomography

Jaewhan Oh[1], Sunggu Kim[2], Chaehwa Jeong[1], Jason Manassa[3], Jonathan Schwartz[3,4], Sangmoon Yoon[5], Robert Hovden[3], Hye Ryung Byon[2†] and Yongsoo Yang[1,6*]

[1]*Department of Physics, Korea Advanced Institute of Science and Technology (KAIST), Daejeon, Republic of Korea*
[2]*Department of Chemistry, Korea Advanced Institute of Science and Technology (KAIST), Daejeon, Republic of Korea*
[3]*Department of Materials Science and Engineering, University of Michigan, Ann Arbor, MI, USA*
[4]*Chan Zuckerberg Imaging Institute, Redwood City, CA, USA*
[5]*Department of Physics and Semiconductor Science, Gachon University, Seongnam, South Korea*
[6]*Graduate School of Semiconductor Technology, School of Electrical Engineering, Korea Advanced Institute of Science and Technology (KAIST), Daejeon, Republic of Korea*

*Corresponding author, email: †hrbyon@kaist.ac.kr, *yongsoo.yang@kaist.ac.kr

**Understanding degradation in battery cathodes and other functional materials requires simultaneous knowledge of structural, chemical, and electronic changes in three dimensions (3D). Here, we present a simultaneous ADF-EDS-EELS tomography method that enables 3D mapping of atomic structure, composition, valence states, and transition metal inhomogeneity within a single, low-dose STEM tilt series acquisition. Applied to LiNi$_{1/3}$Co$_{1/3}$Mn$_{1/3}$O$_2$ (NCM111) particles at different electrochemical cycling stages, this method reveals nanoscale degradation processes with full spatial correlation. We observe that while chemical composition evolves uniformly throughout the entire primary particle, valence state changes and transition metal segregation are strongly depth-dependent and concentrated near the surface. This coexistence of bulk and surface-driven degradation dynamics reveals distinct mechanisms acting at different spatial scales. The evolution of inhomogeneity and valence states deviates from simple phase transition models, highlighting the roles of ion migration and dissolution-driven segregation. Our findings establish valence state gradients and nanoscale inhomogeneity as active contributors to cathode failure. More broadly, this correlative 3D platform opens new opportunities for studying redox-driven transformations in fields such as neuromorphic computing and heterogeneous catalysis, where composition-structure-function coupling is inherently three-dimensional.**

# Main

**Introduction**

   Understanding how structural, chemical, and electronic properties evolve in three dimensions (3D) is central to unraveling the functional behavior and degradation of complex materials. In energy storage, catalysis, electronics, and quantum systems alike, correlated changes in local composition, valence states, and crystal structure govern performance and failure. These transformations often proceed heterogeneously along depth gradients, at interfaces, or via nanoscale segregation and remain difficult to resolve using conventional two-dimensional (2D) or single-modality techniques. This limitation has been especially acute in beam-sensitive materials, where repeated or sequential measurements introduce beam-induced damage and alignment errors that hinder truly correlative 3D analysis. A generalizable method for capturing nanoscale 3D maps of structure, chemical composition, and valence states in a single, low-dose acquisition could provide transformative insights across multiple disciplines.

   In lithium-ion batteries, layered cathode materials such as $LiNi_xCo_yMn_{1-x-y}O_2$ (NCM)[1–4] suffer from progressive capacity fading during repeated charge-discharge cycles[5]. This degradation involves interrelated changes in structure, chemical composition, and valence states that evolve across multiple spatial and temporal scales. It has been attributed to a combination of transition metal dissolution[6–8], local composition inhomogeneity[9,10], and oxygen loss[11]. These process collectively drive structural phase transitions[12], microcrack formation[13], and phase transformations[11], ultimately leading to long-term capacity loss[14,15].

   Despite extensive studies on lithium-ion battery cathodes, many key degradation mechanisms remain poorly understood due to the lack of spatially correlated, depth-resolved measurements that capture structure, chemical composition, valence states, and local inhomogeneity simultaneously. For example, phase transitions from layered to spinel or rock-salt structures often occur non-uniformly and are closely coupled with oxygen loss, cation migration, and changes in transition metal valence states[12,16]. Furthermore, local transition metal inhomogeneity such as nanoscale segregation can alter redox behavior, drive disproportionation reactions, or facilitate dissolution, directly impacting capacity retention and cycle life[17].

   Various microscopy and spectroscopy techniques, including annular dark-field scanning transmission electron microscopy (ADF-STEM), energy-dispersive X-ray spectroscopy (EDS), electron energy-loss spectroscopy (EELS), and transmission X-ray microscopy (TXM), have provided valuable insight into the atomic structure, chemical composition, and valence states of NCM materials during cycling[9,18,19]. However, these methods typically rely on 2D projections, making it difficult to resolve depth-dependent features and to correlate structure, chemical composition, and valence states in 3D. Recent developments in tomography-based transmission electron microscopy (TEM) and TXM have enabled 3D imaging of structure or chemical composition[20–23], yet no technique has previously allowed simultaneous, spatially correlated 3D mapping of structure, chemical composition, and valence states at the nanoscale.

In this study, we developed a simultaneous ADF-EDS-EELS tomography approach to achieve full 3D nanoscale mapping of structure, chemical composition, and valence states, revealing distinct degradation behaviors across electrochemical cycling. This method provided several key advantages: it eliminated the need for multiple separate tilt series measurements for each modality, thereby reducing cumulative electron dose and minimizing beam-induced damage. It also ensured precise spatial alignment among ADF, EDS, and EELS data channels without complex post-registration; and it enabled truly correlative analysis across structural, chemical, and electronic information. By applying this technique to $LiNi_{1/3}Co_{1/3}Mn_{1/3}O_2$ (NCM111) primary particles at different electrochemical cycling stages, we quantitatively captured the coupled evolution of chemical composition and valence states in 3D. Our measurements showed that chemical composition evolved uniformly throughout the primary particles. In contrast, valence state changes and transition metal inhomogeneity were strongly surface-localized and depth-dependent, revealing complex couplings among structural transitions, ion migration, and dissolution that give rise to distinct degradation pathways.

## Results

### Functional and structural degradation during electrochemical cycling

To investigate the structural, chemical, and valence state dynamics in NCM111 cathodes, three different half-cells were prepared. Each cell underwent electrochemical cycling for 50, 100, and 200 cycles at a 1C rate (200 mA $g^{-1}$) with an applied voltage range of 2.7 V–4.3 V (vs. $Li^+$/Li, Methods). Figure 1a presents the galvanostatic charge-discharge curves for the different electrochemical cycles. The cell suffered from continuous degradation, losing approximately 34.3% of its initial capacity (162.1 mAh $g^{-1}$) after 50 cycles. By 100 cycles, 52.2% of the initial capacity was lost, but the cell was still functional. However, after 200 cycles, the cell completely lost its capacity to operate. These results suggest that gradual changes in structure, chemical composition, and valence states occur up to 100 cycles, followed by a drastic transformation between 100 and 200 cycles.

To confirm the structural changes, we conducted atomic-resolution ADF-STEM measurements on the NCM111 cathode primary particles after different electrochemical cycles. As shown in Fig. 1b–d, the pristine particles, initially exhibiting a layered structure [$Li(TM)O_2$, where TM denotes transition metal: Mn, Co or Ni][24], underwent a structural phase transition to a spinel-like structure [$Li(TM)_2O_4$] after 100 cycles[25]. By 200 cycles, the material further transformed into a rock-salt structure [$Li(TM)O$][2], consistent with the typical structural degradation process of NCM cathodes[12]. However, 2D structural measurements alone are insufficient to fully capture degradation dynamics because they inherently lack depth resolution, which is essential since near-surface changes critically affect Li$^+$ transport during battery operation. Moreover, structural information alone cannot reveal local chemical composition or valence state distribution, both of which are strongly interrelated with structure and collectively govern battery behavior.

**Simultaneous ADF-EDS-EELS tomography**

We executed simultaneous ADF-EDS-EELS tomography to acquire full 3D information regarding the dynamics of structure, chemical composition, and valence state during electrochemical cycling, as illustrated in Fig. 1e. During the tomography measurements, an ADF image, EDS spectra, and EELS spectra can be simultaneously obtained at each tilt angle. This simultaneous scheme offers two key advantages: (1) it eliminates the need for multiple electron beam exposures to separately acquire ADF, EDS, and EELS tilt series, substantially reducing the electron dose; (2) all data are acquired in a single STEM measurement, ensuring precise spatial alignment across different modalities without requiring a complex registration process.

We applied this scheme to measure NCM111 cathodes under the four cycling conditions: pristine, after 50 cycles, after 100 cycles, and after 200 cycles. A total of 12 primary particles were selected for simultaneous tomography measurements (Methods): 4 pristine (named pristine particle 1–4), 2 after 50 cycles (50-cycled particle 1–2), 4 after 100 cycles (100-cycled particle 1–4), and 2 after 200 cycles (200-cycled particle 1–2).

The 3D structure and chemical composition maps were reconstructed from the ADF and EDS tilt series via the GENFIRE algorithm[26] (Methods). The 3D chemical composition maps were further validated using a fused multi-modal tomography reconstruction[27], which showed consistent distributions across all transition metals, confirming the reliability of our analysis (Supplementary Fig. 1).

Since the allowed electron dose level was severely limited to prevent specimen damage of beam-sensitive NCM cathodes (about three orders of magnitude lower dose compared to typical EELS tomography measurements[21,28,29]), the inelastic EELS signal suffers from an extremely low signal-to-noise ratio (SNR). Therefore, proper estimation of valence states from the EELS tilt series required a noise-tolerant background subtraction pipeline and spectral decomposition process. To address this, we developed an iterative background determination method based on the Hartree–Slater continuum model[30] and implemented a non-negative matrix factorization (NMF) algorithm[31] to selectively extract contributions from different valence states in the spectra (Methods). Final reconstructed 3D valence state maps for Mn ($Mn^{4+}$, $Mn^{3+}$, $Mn^{2+}$) and Co ($Co^{3+}$, $Co^{2+}$) were reconstructed from the extracted tilt series of 2D valence state maps (Methods). Note that analyzing the $Co^{4+}$ valence state via EELS is challenging due to the absence of a stable reference sample[32–35]. Additionally, we could not determine the valence states of Ni due to the low SNR near the Ni L edge, which prevented reliable distinction of their spatial distributions.

**Evolution of 3D chemical composition of NCM111**

The EDS results revealed the 3D chemical distribution of oxygen and transition metals across different electrochemical cycles (Fig. 2a–d, Supplementary Fig. 2, and Supplementary Movie 1). A clear change in the distribution of oxygen and transition metals can be qualitatively observed from the reconstructed volumes, particularly between 100 and 200 cycles, where a sudden decrease in oxygen and an increase in transition metals occur. This change is consistent with the expected transition from the layered or spinel-like structure to the rock-salt structure (Fig. 1b–e)[12,36].

The pristine, 50-cycled, and 100-cycled particles exhibit a homogeneous distribution of oxygen and transition metals. However, small regions with higher concentrations of transition metals, indicating local segregation, are visible, particularly near the surface of the pristine particle (e.g., the regions highlighted by the blue circles in Fig. 2a–d). In contrast, the 200-cycled particle exhibits a notably inhomogeneous distribution of transition metals, pointing to substantial chemical composition changes occurring between 100 and 200 cycles. Beyond qualitative observations, our chemical composition maps enable a rigorous quantitative analysis of the chemical evolution of NCM111 cathodes. First, we calculated the average atomic ratios of O, Mn, Co, and Ni from the 3D chemical composition maps of the entire particle (Fig. 2e). The average atomic percentage of oxygen remained nearly constant up to 100 cycles (66.0% to 67.0%) but showed a sharp decline at 200 cycles (dropping to 61.0%). Although a gradual transition from the layered to the spinel-like structure is expected up to 100 cycles, our composition mapping cannot detect this transition, as both structures share the same 1:2 transition metal-to-oxygen ratio. However, by comparing the transition metal-to-oxygen atomic percentage ratio of the layered or spinel-like structures (66.7%) with that of the rock-salt structure (50.0%), we quantitatively determined that 33.7% of the particle volume has transformed into the rock-salt structure after 200 cycles.

Second, we quantified transition metal segregation by calculating the percentage of oxygen and transition-metal-rich regions within the entire volume of the measured particles, as described in Fig. 2f (Methods). Up to 50 cycles, oxygen and transition-metals remained uniformly distributed throughout the particles, with an almost negligible fraction of transition metal-rich regions. By 100 cycles, a small fraction (0.1%) of transition-metal-rich regions emerged, and by 200 cycles, this fraction increased to approximately 0.2–0.5%.

One of the key advantages of tomography over 2D projection-based analysis is its ability to resolve depth information. Using the 3D volume data, we segmented the surface regions and compared their behavior with that of the entire volume during cycling (Fig. 2g,h). Notably, the evolution of oxygen and transition metal atomic percentages at the surface closely mirrors the trend observed for the full volume: oxygen content remains stable up to 100 cycles and then begins to decline (Fig. 2g). Likewise, the increase in the fraction of transition-metal-rich regions is consistently observed in both the surface and the entire volume (Fig. 2f,h). This parallel behavior is somewhat unexpected, as cycling-induced changes are typically expected to occur predominantly at the surface and along intergranular microcracks, especially at the scale of secondary particles[18,37,38].

To investigate further, we quantitatively analyzed the depth profiles of O, Mn, Co, and Ni atomic percentages. Depth profiles were extracted from individual particles, and those from particles that underwent the same number of cycles were averaged to produce mean depth profiles for each cycling condition, as shown in Fig. 2i–l. The error bars, generally within 1%, indicate a strong consistency of chemical composition across particles with the same cycling history. A comparison of the depth profiles reveals that the pristine, 50-cycled, and 100-cycled particles show similar atomic percentages, whereas the 200-cycled particles display a distinct compositional shift throughout the profile. This result aligns with our earlier observations (Fig. 2e,g), showing minimal chemical composition changes up to 100 cycles, and a pronounced shift at 200 cycles.

Notably, the chemical composition remains largely uniform throughout the particles, regardless of the depth (Fig. 2i–l). The only exception is the profiles of 100-cycled particles, which exhibit depth dependence. In fact, individual depth profiles of 100-cycled particles before averaging show a uniform profile (Supplementary Fig. 3), whereas averaging introduces apparent depth dependence with large error bars due to compositional variations among different particles, particularly for Co. Thus, our analysis confirms that chemical composition is generally uniform across all measured particles, with no intrinsic depth dependence. This finding supports recent reports that structural transitions extend beyond the surface region and occur throughout the entire primary particle, highlighting the necessity of depth-sensitive 3D measurements to fully capture degradation mechanisms[17,39,40].

Unlike the compositional changes, which occur relatively uniformly throughout the primary particles, transition metal segregation is substantially more pronounced near the surface. As shown in Fig. 2f,h, the degree of segregation at 100 and 200 cycles is approximately twice as high in the surface region compared to the entire volume. Among the transition metals, Co exhibits the most pronounced increase in the fraction of transition-metal-rich regions (i.e., increased local segregation), followed by Ni, while Mn shows the least. This trend is consistent with the fact that transition metal dissolution during electrochemical cycling primarily occurs at the surface[41], and aligns with the known dissolution tendencies of each element: Co is more resistant to dissolution and thus more likely to accumulate, whereas Mn dissolves faster and is less likely to remain as rich regions[41].

These findings reveal two coexisting degradation processes during cycling: (1) uniform compositional changes affecting both surface and interior regions, and (2) local transition metal segregation that is more prominent at the surface. This segregation is closely linked to dissolution behavior, which is influenced not only by chemical composition but also critically by the valence states of the transition metals[41–43]. Therefore, to fully decode the underlying mechanisms, particularly those governing local transition metal segregation, 3D valence state information is required in addition to structural (ADF) and chemical (EDS) analyses.

**Evolution of 3D valence state of transition metals**

Our 3D valence state maps provide important clues regarding the relationship between transition metal dissolution and structural phase transitions, particularly those that occur without accompanying compositional changes, such as the transition from the layered to the spinel-like structure. Even before full 3D reconstruction, simple 2D analysis from our EELS tilt series reveals clear valence state changes during cycling, suggesting valuable information is embedded in our data (Supplementary Fig. 4).

A qualitative overview of the reconstructed 3D valence state maps shows that $Mn^{4+}$ and $Co^{3+}$ dominate throughout the particle in the pristine state, consistent with the expected electronic structure of the layered NCM111 cathode (Fig. 3a–e, Supplementary Fig. 5, and Supplementary Movies 2 and 3)[44–46]. While the proportions of these valence states remain relatively stable up to 100 cycles, a substantial decrease is observed at 200 cycles, along with the emergence of lower valence states. Quantitative analysis reveals a considerable drop in the fractions of $Mn^{4+}$ (from 80.7% to 53.4%) and $Co^{3+}$ (from 77.5% to 11.0%), accompanied by corresponding increases in $Mn^{2+}$ (from 12.1% to 42.0%) and $Co^{2+}$ (from 22.5% to 89.0%) between 100 and 200 cycles (Fig. 3f,g). These valence shifts are indicative of structural phase transitions during cycling; a reduction of valence state (from $Mn^{4+}$ to $Mn^{2+}$ and $Co^{3+}$ to $Co^{2+}$) is characteristic of the transition from the layered ($LiTMO_2$) to the rock-salt (LiTMO) structure[12], which is also supported by our ADF-STEM imaging and compositional analysis.

The evolution of valence state percentages for both Mn and Co show a consistent trend in both the entire volume (Fig. 3f,g) and the surface region (Fig. 3h,i), though the absolute values differ. This observation supports that electrochemical reactions driving structural and chemical transformations occur not only at the surface but also throughout the entire primary particle, consistent with our earlier argument that distinct degradation mechanisms may coexist: one affecting the whole particle uniformly and another localized near the surface.

**Depth-dependent valence states and the role of local segregation**

We further investigated the depth-dependent profiles of Mn and Co valence states based on our 3D valence state maps (Fig. 3j–n and Supplementary Fig. 6,7). In sharp contrast to the uniform depth profiles observed in chemical composition (Fig. 2i–l), the valence states of transition metals exhibit strong depth dependence, which is likely related to transition metal segregation and dissolution, as hinted by our EDS analysis.

Even in the pristine state, a pronounced depth dependence in valence states is evident. The $Mn^{4+}$ fraction increases with depth (Fig. 3j), while $Mn^{3+}$ and $Mn^{2+}$ exhibits the opposite trend (Fig. 3k,l), and this behavior persists up to 100 cycles. After 200 cycles, however, $Mn^{4+}$ decreases while $Mn^{3+}$ and $Mn^{2+}$ increase toward the core. A similar trend is observed for Co: up to 100 cycles, $Co^{3+}$ increases and $Co^{2+}$ decreases with depth, but this trend reverses after 200 cycles (Fig. 3m,n).

The core regions of the pristine particles exhibit valence states consistent with the ideal layered NCM111 structure ($Mn^{4+}$ and $Co^{3+}$)[12,45,46], while deviations emerge toward the surface. This depth profile may arise from several possible scenarios: (1) depth-dependent oxygen vacancies (which lower valence states), (2) excess lithium with depth-dependent Li concentration, (3) formation of secondary structural phase(such as spinel or rock-salt formation), or (4) element-specific local segregation of individual transition metals, which can alter bonding and charge balance while preserving the layered structure[47].

We examined each possibility one by one. Our 3D chemical composition maps (Fig. 2i and Supplementary Fig. 3) reveal a uniform oxygen distribution, ruling out depth-dependent oxygen vacancy gradients. Inductively coupled plasma optical emission spectrometry (ICP-OES) of pristine NCM111 shows no evidence of excess lithium (Supplementary Table 1), eliminating lithium enrichment as a contributing factor. Furthermore, powder X-ray diffraction (XRD) and atomic-resolution ADF-STEM confirm a well-ordered layered structure with no signs of spinel or rock-salt phases (Fig. 1b–d and Supplementary Fig. 8a,b), thereby excluding secondary phase formation.

The remaining explanation is local segregation of individual transition metals near the surface. Such segregation can distort local bonding environments and disrupt charge balance, giving rise to lower valence states such as $Mn^{3+}$ and $Co^{2+}$ (ref. [47]). Evidence of this segregation is clearly seen in our 3D chemical composition maps (Fig. 2a–d and Supplementary Fig. 2), which show surface-localized clustering of Mn, Co, and Ni in pristine NCM111.

To further explore this effect, we quantified the depth-dependent inhomogeneity of each transition metal using EDS tomography (Fig. 4a,b), where inhomogeneity is defined as the standard deviation of the chemical composition at a given depth (Methods). The results show that Mn and Co inhomogeneity is highest at the surface and gradually decreases toward the core, confirming that transition metal segregation is more prominent near the particle surface.

Taken together, our structural, chemical, and valence state analyses suggest that the pristine particles maintain the layered structure but exhibit surface-localized clustering of individual transition metals, which leads to corresponding variations in their local valence states[47].

**Degradation dynamics at different stages of battery cycling**

Our results reveal a strong correlation between transition metal inhomogeneity (localized segregation of specific elements) and valence states. This interplay influences their evolution during electrochemical cycling and plays a critical role in structural degradation, particularly near the surface[18,48]. To gain deeper insight into these degradation dynamics, we analyzed the coupled behavior of Mn inhomogeneity and valence states.

As shown in Fig. 4a, the evolution of Mn inhomogeneity exhibits two distinct features. First, across all cycling conditions, Mn inhomogeneity consistently decreases with depth, following an exponential-like decay profile. This indicates that both its formation and evolution are depth-dependent and primarily concentrated near the surface of the primary particles. This trend aligns with the depth profiles of Mn valence states observed in Fig. 3j–l. Second, the inhomogeneity profile shows non-monotonic cycling evolution. Compared to the pristine state, Mn inhomogeneity decreases after 50 cycles, suggesting a more uniform Mn distribution (i.e., reduced segregation). However, it then gradually increases between 50 and 200 cycles. A similar non-monotonic trend is observed in the Mn valence states (Fig. 3j–l); between the pristine and 50-cycle states, $Mn^{4+}$ content decreases by approximately 10%, accompanied by an increase in $Mn^{3+}$. Between 50 and 100 cycles, this trend reverses: $Mn^{4+}$ increases, $Mn^{3+}$ decreases, and $Mn^{2+}$ slightly declines. By 200 cycles, $Mn^{2+}$ content rises sharply, reaching around 40%.

The valence states and inhomogeneity of Mn are generally expected to be correlated, as lower valence states such as $Mn^{3+}$ and $Mn^{2+}$ are anticipated to appear in regions where Mn segregation occurs[47]. In this case, increasing inhomogeneity should correspond to a decrease in $Mn^{4+}$ content, and vice versa. However, the observed trends deviate from this expected correlation. Up to 100 cycles, both inhomogeneity and $Mn^{4+}$ increase or decrease simultaneously, in contrast to the anticipated inverse relationship. This deviation suggests that the evolution of Mn valence states cannot be fully explained by structural phase transitions or segregation alone. Instead, it highlights the need to consider the non-conserved nature of transition metals, particularly the loss of metal ions through chemical reactions occurring at different stages of battery cycling. To further clarify this behavior, we analyze each stage separately.

During the first 50 cycles, both $Mn^{4+}$ content and Mn inhomogeneity decrease (Fig. 3f,j–l and Fig. 4a). This behavior can be explained with a structural phase transition and accompanying transition metal migration. Specifically, the transformation from the layered structure ($LiTMO_2$, $Mn^{4+}$) to the spinel-like structure ($LiTM_2O_4$, containing $Mn^{4+}$ and $Mn^{3+}$) introduces $Mn^{3+}$ into the system, thereby reducing the overall $Mn^{4+}$ fraction. Concurrently, transition metal ions (particularly Ni, which has the lowest migration energy barrier[49]) migrate into the lithium layers. This redistribution of transition metals is expected to reduce spatial variations in local transition metal concentration, leading to a more homogeneous distribution[50,51] and, consequently, lower Mn inhomogeneity.

Beyond 50 cycles, the trend reverses: both the fraction of $Mn^{4+}$ and Mn inhomogeneity increase up to 100 cycles (Fig. 3f,j–l and Fig. 4a). This behavior cannot be explained by structural transitions alone. At this stage, the growing presence of spinel-like regions and the accompanying increase in $Mn^{3+}$ content suggest that the disproportionation reaction must be considered. This reaction, triggered by interactions with HF in the electrolyte, results in the formation of insoluble $Mn^{4+}$ and soluble $Mn^{2+}$, leading to a net loss of Mn from the structure due to dissolution[42,43,52]. As the reaction progresses, it promotes the formation of Mn-rich secondary phases such as segregated $MnO_2$ or $MnF_3$ (ref.[43,53,54]). These localized, high-Mn regions are clearly visible in our 3D chemical composition maps and directly contribute to the increased spatial variance and the observed rise in Mn inhomogeneity.

Between 100 and 200 cycles, the structure evolves further toward the rock-salt phase (LiTMO, $Mn^{2+}$), which is consistent with the sharp increase in $Mn^{2+}$ content observed during this stage (Fig. 3f,j–l).

To determine whether similar behavior occurs for other transition metals, we also investigated the correlation between Co inhomogeneity and valence states. $Co^{3+}$ content remains relatively stable up to 100 cycles but decreases sharply at 200 cycles, with a corresponding rise in $Co^{2+}$ (Fig. 3g,m,n). Co inhomogeneity follows a similar pattern to Mn: it decreases between the pristine and 50-cycle states, then increases steadily through 200 cycles (Fig. 4b). The initial decrease is again attributed to transition metal migration during structural transition, which reduces spatial variance in Co content. After 50 cycles, the formation of segregated surface phases such as $Co_3O_4$ and $CoF_3$ becomes dominant[53–55], driving the increase in Co inhomogeneity. This progression closely mirrors the trend observed for Mn.

Our findings indicate that three key mechanisms occur during battery cycling: (1) structural phase transitions, (2) transition metal migration (which reduces inhomogeneity even in the absence of structural change) and (3) transition metal segregation. The interplay among these processes governs the evolution of transition metal valence states, which directly impacts electrochemical performance and long-term durability[17]. Importantly, our results highlight that transition metal inhomogeneity, valence states, and their depth-dependent profiles within primary particles are not merely indicators of degradation but may also serve as tunable parameters for improving the stability and lifetime of NCM cathode materials.

**Conclusion**

In this work, we achieve comprehensive 3D mapping of structure, chemical composition, valence states, and transition metal inhomogeneity in battery cathode materials through a simultaneous ADF-EDS-EELS tomography framework. Applied to NCM111 cathodes across distinct electrochemical cycling stages, our approach elucidates nanoscale degradation phenomena with full spatial and chemical fidelity.

Our measurements revealed that localized transition metal inhomogeneity is already present in pristine particles and becomes increasingly concentrated near the surface during cycling. These inhomogeneities were strongly coupled to depth-dependent shifts in valence states. However, the evolution of these parameters did not follow simple or monotonic trends, revealing that structural phase transitions alone cannot fully explain degradation. Instead, our results point to a complex interplay among structural phase transitions, transition metal migration, and dissolution-driven segregation as key degradation pathways.

Importantly, we observed that compositional changes occurred uniformly throughout the particle volume, whereas both valence state evolution and transition metal inhomogeneity were dominated by strong surface-depth gradients. These results highlight that 3D valence state distribution and chemical inhomogeneity serve not only as signatures of degradation, but also as active contributors that determine the functional stability and long-term performance of NCM cathodes.

More broadly, our simultaneous ADF-EDS-EELS tomography platform provides a powerful and generalizable method for decoding correlated structural, chemical, and electronic behaviors in complex materials. Because these properties and their interplay are fundamental to a wide range of functional materials and devices, the method is broadly applicable beyond battery research. For instance, understanding redox-driven structural distortions in transition metal oxides for neuromorphic computing or unraveling catalytic site evolution in heterogeneous catalysts for sustainable fuel production both require precisely this kind of spatially resolved, multimodal analysis.

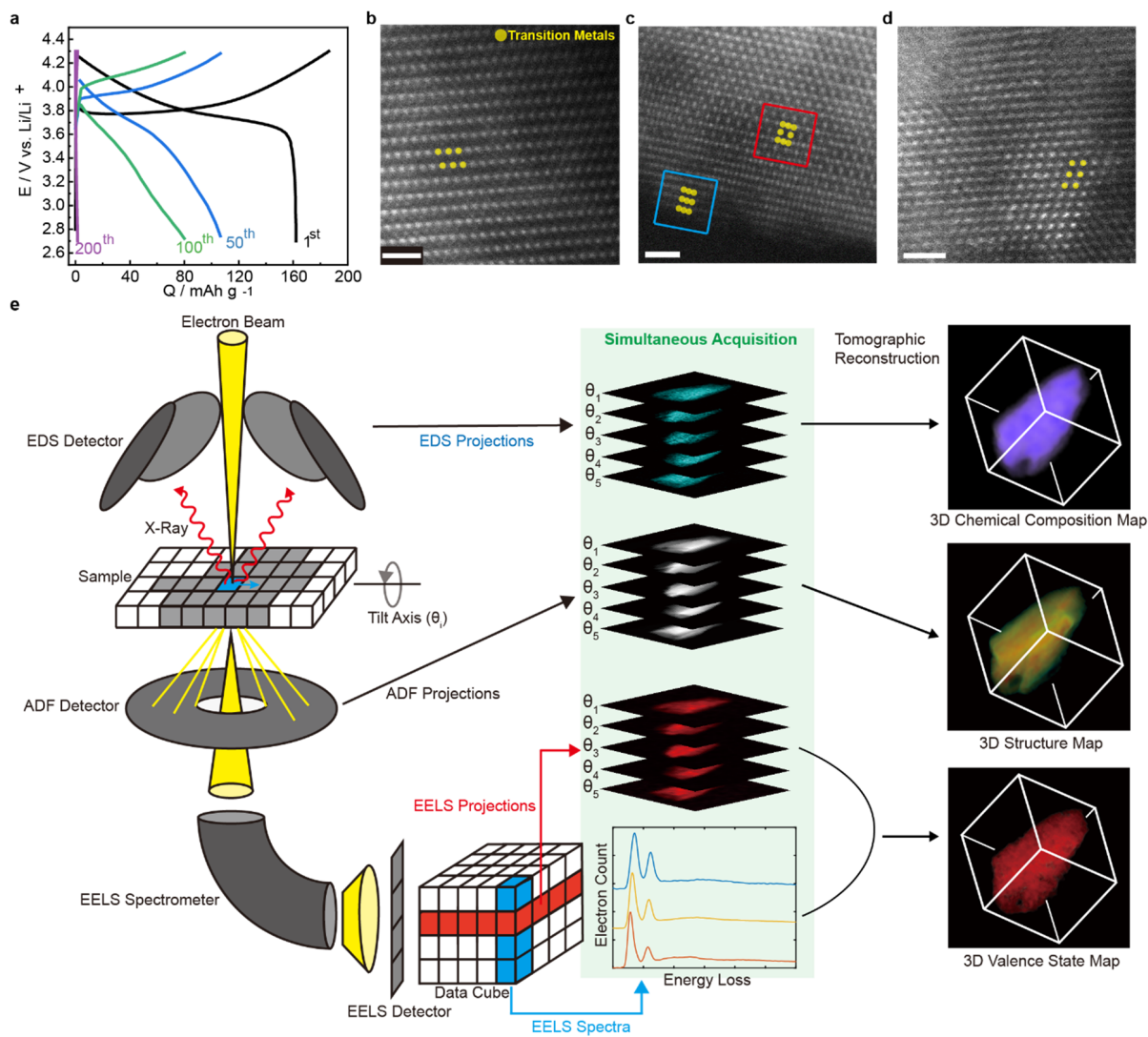

**Figure 1 | Electrochemical characteristics, change in lattice structure, and simultaneous ADF-EDS-EELS tomography workflow. a,** Charge–discharge curves of NCM111 cathodes at the 1st, 50th, 100th, and 200th cycles. **b–d,** Atomic-resolution ADF-STEM images of NCM111 primary particles in the pristine state **(b)**, after 100 cycles **(c)**, and after 200 cycles **(d)**, revealing structural phase transitions from the original layered Li(TM)$O_2$ structure to the spinel-like Li(TM)$_2$$O_4$ structure and eventually to the rock-salt Li(TM)O structure. Yellow circles indicate representative transition metal atomic columns and red and blue boxes in **(c)** highlight the spinel-like and rock-salt structural regions, respectively. Scale bar, 1 nm. **e,** Schematic of simultaneous ADF-EDS-EELS tomography workflow. A single electron tomography acquisition enables reconstruction of 3D structure, chemical composition, and valence state maps.



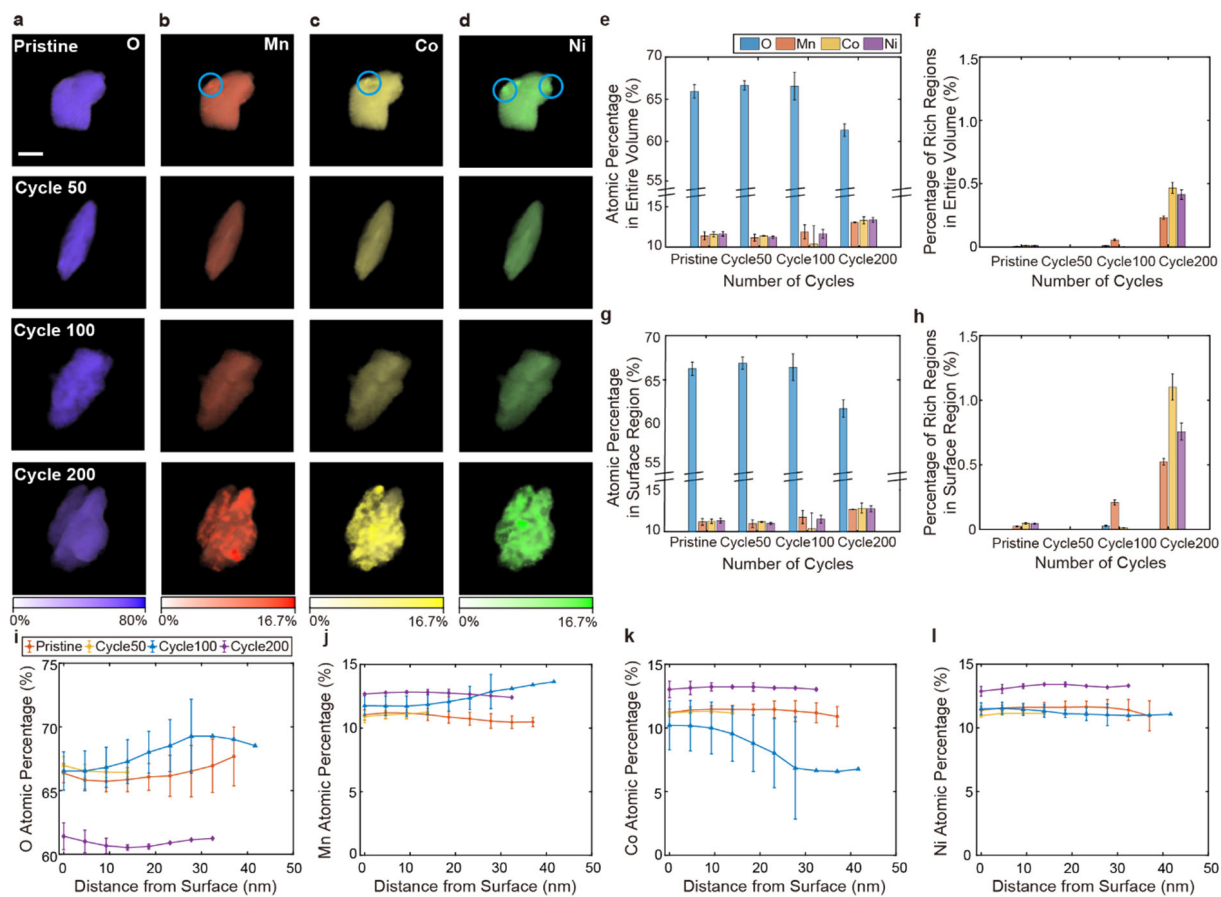

**Figure 2 | Chemical evolution and depth profiles of oxygen and transition metals during electrochemical cycling. a–d**, 3D rendering of chemical composition maps of particle 2 from the pristine samples (1st row), particle 1 from the 50-cycled samples (2nd row), particle 1 from the 100-cycled samples (3rd row), and particle 1 from the 200-cycled samples (4th row) for O (**a**), Mn (**b**), Co (**c**), and Ni (**d**), respectively. Scale bar, 50 nm. **e, f**, Changes in chemical composition (**e**) and the percentages of oxygen and transition-metal-rich regions (**f**) during electrochemical cycling for the entire particle. Note that these percentages were averaged over the values from particles that experienced the same number of cycles. **g, h** Similar plot with (**e,f**) for surface region only: chemical composition (**g**) and fraction of oxygen- and transition-metal-rich regions (**h**) during electrochemical cycling, analyzed using the same averaging method. **i-l**, Evolution in depth profiles of chemical composition for O (**i**), Mn (**j**), Co (**k**), and Ni (**l**), respectively. These profiles were obtained by averaging the data from multiple particles that experienced the same cycling condition. All error bars represent standard deviations at each depth interval, indicating variation in atomic percentage at the corresponding depth.



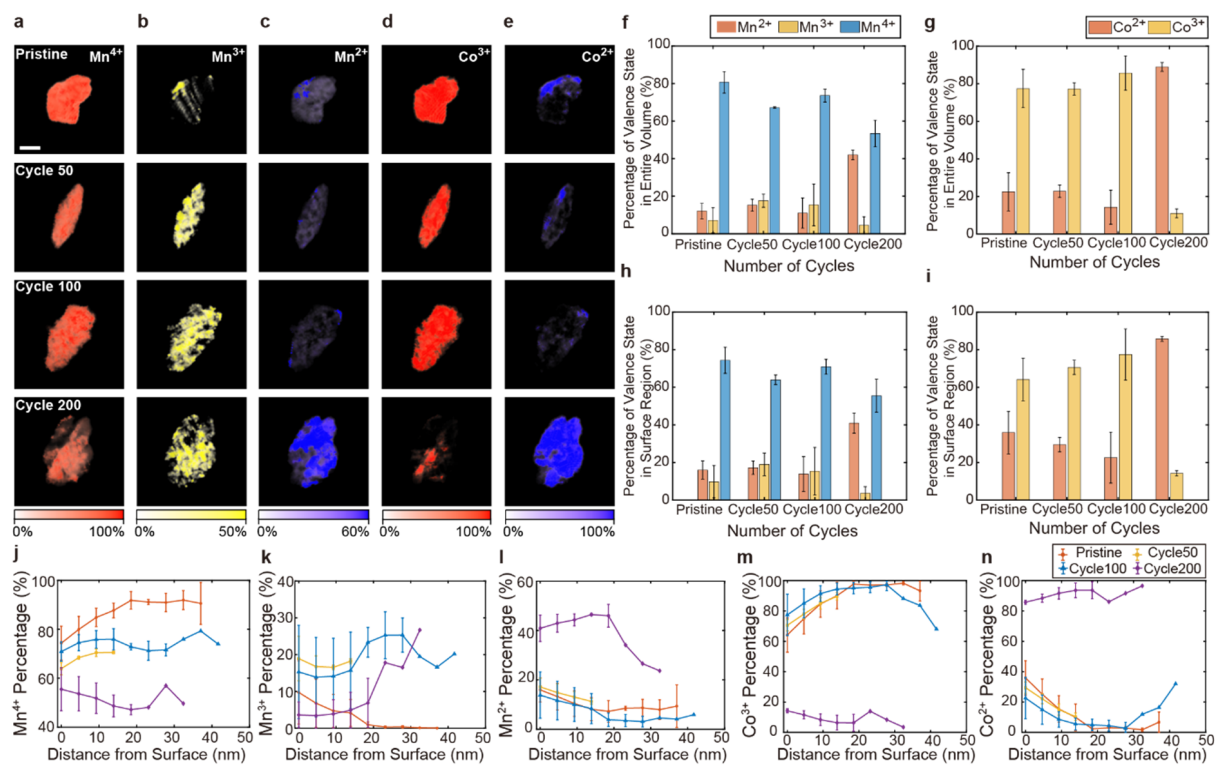

**Figure 3 | Evolution of Mn and Co valence states and depth profiles during electrochemical cycling. a–e**, 3D rendering of Mn and Co valence state maps at each cycling condition, from the same particles shown in Fig. 2a–d, for $Mn^{4+}$ (**a**), $Mn^{3+}$ (**b**), $Mn^{2+}$ (**c**), $Co^{3+}$ (**d**), and $Co^{2+}$ (**e**). Scale bar, 50 nm. **f, g,** Evolution of Mn (**f**) and Co (**g**) valence states during electrochemical cycling for the entire particle. **h, i,** Evolution of Mn (**h**) and Co (**i**) valence states during electrochemical cycling for the surface region. **j–n,** Evolution in depth profiles of valence states for $Mn^{4+}$ (**j**), $Mn^{3+}$ (**k**), $Mn^{2+}$ (**l**), $Co^{3+}$ (**m**), and $Co^{2+}$ (**n**), respectively. All values and profiles in this figure were obtained using the same averaging procedure described in Fig. 2. All error bars represent standard deviations at each depth interval, indicating variation in the percentage of valence states at the corresponding depth.



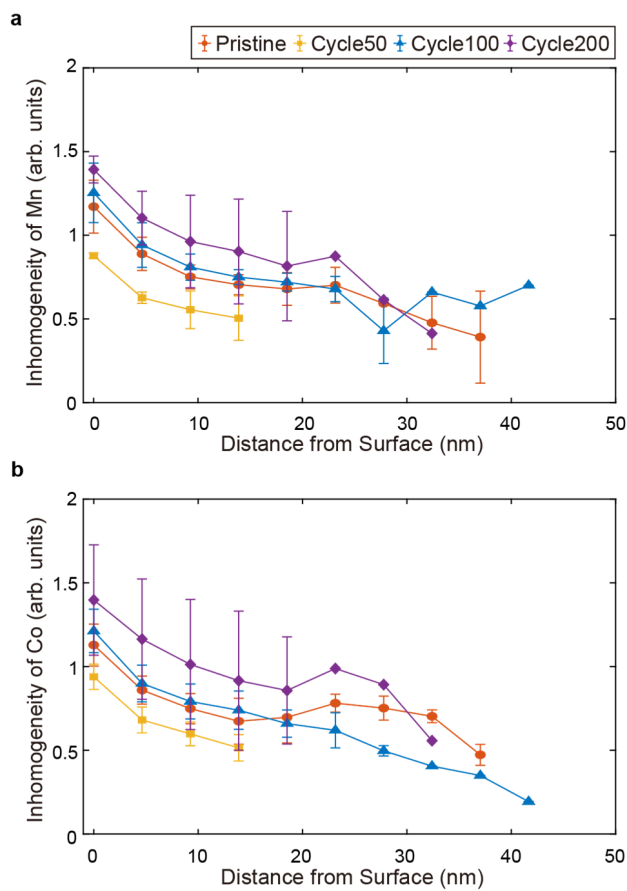

**Figure 4 | Depth profiles of chemical inhomogeneity. a, b,** Depth profiles of chemical inhomogeneity for Mn **(a)** and Co **(b)** during electrochemical cycling. These profiles were obtained by averaging the inhomogeneity depth profiles of multiple particles under the same cycling condition. All error bars represent standard deviations at each depth interval.



# METHODS

## Initial characterization

NCM111 samples were purchased from Sigma-Aldrich. The atomic structure and chemical composition of the NCM111 pristine particles were initially characterized using powder XRD and ICP-OES. Powder XRD measurements were conducted using a Rigaku SmartLab equipped with a Cu Kα radiation source (λ = 1.54 Å, 8.04 keV) at the KAIST Analysis Center for Research Advancement (KARA). The diffraction pattern was collected over a 2θ range of 10.0° to 80.0° with a step size of 0.01° (Supplementary Fig. 8a). The obtained XRD pattern was compared to the standard reference card (00-062-0431), confirming that the sample exhibits the layered structure (Supplementary Fig. 8b).

Elemental quantification was carried out using an Agilent ICP-OES 720 system to determine the chemical composition of Li, Mn, Co, and Ni in the pristine particles. The measured elemental ratios revealed a Li:Mn:Co:Ni stoichiometry of approximately 6:1:1:1 (Supplementary Table 1), in agreement with the expected chemical composition of the NCM111 cathode.

## Half-cell setup and electrochemical testing

A 0.2 g batch of cathode slurry was prepared by mixing 80 wt% NCM111 powder, 10 wt% conductive carbon (Super P, Timcal), and 10 wt% polyvinylidene fluoride (PVDF, Kynar HSV900) in N-methyl-2-pyrrolidone (NMP, Sigma-Aldrich, 99%). The components were mixed using a planetary mixer (ARE-310, THINKY corporation) at 1000 rpm for 20 minutes. The resulting slurry was cast onto aluminum foil (MTI, purity > 99.3%) using a micrometer film adjustable applicator (Wellcos), and subsequently dried in an oven at 80 °C. The loading weight of the NCM111 electrode was in range of 2–3 mg cm$^{-2}$. The electrolyte used was 1.0 M LiPF$_6$ in a 1:1 volume ratio of ethylene carbonate (EC) and diethyl carbonate (DEC). A lithium metal disk (Honjo, 15 mm diameter, 150 μm thickness) and a Celgard2500 membrane were used as the counter electrode and separator, respectively. Half-cells were assembled in CR2032 coin cells (Wellcos) inside an argon-filled glove box (MOTEK) with oxygen and moisture levels maintained below 1.0 ppm, and ~0 ppm, respectively.

Electrochemical testing was performed by charging and discharging each coin cell under galvanostatic conditions at a rate of 1C (200 mA g$^{-1}$) between 2.7 V and 4.3 V (vs Li$^+$/Li) using a temperature-controlled battery cycler (PNE) operated at room temperature.

## TEM sample preparation

Specimens were prepared using NCM cathodes under four different electrochemical cycling conditions: pristine (0-cycles), 50-cycles, 100-cycles, and 200-cycles. After electrochemical cycling, each NCM111 powder sample was dispersed in ethanol (purity > 99.3%) and sonicated for 30 minutes.



Then, 3.50 μL of the dispersed solution was drop-cast onto a 4 nm thick ultrathin carbon grid. The prepared grids were heat-treated at 150 °C under low vacuum conditions for 8 hours to minimize hydrocarbon contamination during data acquisition.

**ADF-STEM measurement**

Atomic-resolution ADF-STEM images were acquired at 300 kV using a double Cs-corrected TEM (Spectra Ultra, Thermo Fisher) at KARA. The images were acquired in ADF-STEM mode using a convergence semi-angle of 21.4 mrad and detector inner and outer semi-angles of 39 mrad and 200 mrad, respectively. The screen current ranged from 56 pA to 77 pA. ADF images were recorded with a scan size of 1024 × 1024 pixels, and the pixel sizes were 6.6 pm, 8.4 pm, and 12.2 pm for the pristine, 100-cycled and 200-cycled samples, respectively.

**Tomographic tilt series acquisition**

Tomographic tilt series were acquired using a Titan double Cs-corrected transmission electron microscope (Titan cubed G2 60-300, FEI) operated at 300 kV. A total of 12 datasets were obtained, consisting of four tilt series each from the pristine and 100-cycled samples, and two each from the 50-cycled and 200-cycled samples.

During each acquisition, ADF, EDS, and EELS signals were simultaneously collected over a tilt range of –70.0° to +70.0° with 10.0° steps. Scanning was performed with a screen current of 0.300 nA, a convergence semi-angle of 25.2 mrad, and a pixel dwell time of 0.05 s. The ADF detector inner and outer semi-angles were 100.0 mrad and 200.0 mrad, respectively. EDS data were acquired using a Super X-detector, and EELS data were collected in dual-EELS mode using a GIF Quantum 965 spectrometer, with a collection semi-angle of 100.0 mrad and an energy dispersion of 0.15 eV. The low-loss and core-loss spectra were collected over the energy ranges of –30.00 eV to 277.05 eV and 600.00 eV to 907.05 eV, respectively.

Each simultaneous ADF-EDS-EELS tomography dataset was acquired with 64 × 64 probe scan positions and a scanning step size of 4.63 nm, resulting in ADF images of 64 × 64 pixels, EDS data of 64 × 64 × 4096 elements, and EELS data of 64 × 64 × 2048 elements per tilt angle. The total electron dose for each dataset was approximately $6.56 \times 10^5$ $e$ Å$^{-2}$.

To evaluate potential electron beam-induced structural changes, zero-degree tilt images were acquired three times: at the beginning, in the middle, and at the end of each tilt series. As shown in Supplementary Fig. 9a–c, no significant structural changes were observed throughout the acquisition. In addition, the zero-degree EELS Mn L$_3$ and Co L$_3$ edge intensity maps were extracted from the corresponding EELS spectra by integrating the signal over the Mn and Co L$_3$ edges, respectively. The resulting maps (Supplementary Fig. 9d–f for Mn and 9g–i for Co) exhibited no noticeable changes, confirming that the valence states of Mn and Co remained stable during the tilt series acquisition.



**ADF and EDS tomographic reconstructions**

1) ADF tomography

Image-post processing was performed on the ADF tilt series before tomographic reconstruction, including binary masking and tilt-series alignment. The Otsu's method[56] was used to generate binary masks for each ADF image, where pixels outside the mask were set to zero. After masking, each tilt series was aligned using the center of mass[57] and common-line alignment[58] methods, following procedures described in previous works[20,59–62]. The aligned ADF tilt series were reconstructed into 3D tomograms using the GENFIRE algorithm[26] with the following parameters: oversampling ratio of 3, interpolation radius of 0.1 pixels, number of iterations of 1000, and discrete Fourier transform interpolation as the interpolation method.

2) EDS tomography

2D elemental intensity maps were extracted from the EDS spectral data cubes using VELOX software, based on the O-$K_\alpha$, Mn-$L_\alpha$, Co-$L_\alpha$, and Ni-$L_\alpha$ characteristic lines (Supplementary Fig. 10). For masking and tilt-series alignment, the same binary masks and alignment parameters used for the ADF tilt series were applied to the EDS tilt series for each element.

To correct for tilt-dependent X-ray absorption and shadowing effects, which arise from the specimen geometry and detector positioning[23,63], a tilt-angle dependent normalization was performed. Specifically, the total sum of pixel intensities of each elemental intensity map was calculated for each tilt angle. Normalization factors were then computed to ensure that this total sum remained constant across all tilt angles. These factors were initially computed for each element (Supplementary Fig. 11a), and final normalization values were obtained by averaging the factors of different chemical elements (O, Mn, Co, and Ni) at each tilt angle (Supplementary Fig. 11b).

After applying the normalization factor, a Gaussian convolution with a standard deviation of 1.2 pixels was applied to the elemental intensity maps, which enhances the SNR with a small trade-off in spatial resolution. The processed EDS intensity tilt series were then reconstructed into 3D intensity tomograms for each element using the same GENFIRE reconstruction parameters used for the ADF reconstructions.

To convert the EDS intensity tomograms into quantitative 3D chemical composition maps, $k$-factors[64] were determined by assuming an ideal composition of O:Mn:Co:Ni = 6:1:1:1, corresponding to the expected stoichiometry of the pristine NCM111 material. These $k$-factors were calculated using the EDS intensity tomograms of four different pristine particles. Using the obtained $k$-factors, the EDS intensity tomograms were converted into 3D chemical composition maps for each element.

**EELS data post-processing and tomographic reconstructions**

1) EELS reference measurement



To analyze the valence states of transition metals in NCM111, EELS reference spectra were acquired. For Mn, reference spectra corresponding to $Mn^{4+}$, $Mn^{3+}$, and $Mn^{2+}$ were collected from $MnO_2$, $Mn_2O_3$, and MnO powder samples, respectively. The spectra were collected over the energy range from 480.00 eV to 684.70 eV with an energy dispersion of 0.10 eV (Supplementary Fig. 12a). Similarly, Co reference spectra for $Co^{3+}$ and $Co^{2+}$ were acquired from $LiCoO_2$ and CoO powder samples, respectively, and collected over the energy range from 630.00 eV to 834.70 eV with the same energy dispersion of 0.10 eV (Supplementary Fig. 12b). All powder reference samples were purchased from Sigma-Aldrich, and deposited onto ultrathin carbon grids for EELS acquisition.

2) EELS drift correction and background subtraction

Energy drift correction was applied to both low-loss and core-loss EELS datasets by aligning the zero-loss peak position in the low-loss spectra. To enhance the SNR in the spatial domain, 3 × 3 pixel (14.5 nm × 14.5 nm) square kernel was applied as a convolution mask across the spectral image stack, equivalent to a mean filter.

Accurate background subtraction is essential for extracting physically meaningful information from the core-loss EELS spectra. We implemented an iterative background subtraction procedure based on two key physical assumptions: (i) the slope of pre-edge region should approach zero after background subtraction, and (ii) the slope of the normalized continuum region in the post-edge should match the theoretical Hartree-Slater continuum model[30,65]. Based on those assumptions, background subtraction was performed using the following steps for selected spectra near the Mn $L_{32}$ edge (600.00 eV – 766.65 eV), and the Co $L_{32}$ edge (718.50 eV – 907.05 eV).

a) Two fitting windows were defined: one in the pre-edge region (pre-edge window) and another in the post-edge region (post-edge window). The post-edge window was positioned more than 70.0 eV beyond the edge onset, where the Hartree–Slater continuum can be approximated as a linear function.

b) Two background model functions were applied to the raw selected spectrum to generate two separate background-subtracted spectra. Background fitting function 1 used only the pre-edge window as a fitting region and employed a power law model[66], defined as

$$Function\ 1)\quad ax^{-b} + c,$$

where *x* is the energy loss, and *a, b, c* are fitting parameters.

Background fitting function 2 used both the pre-edge and post-edge windows as a fitting region, and employed an extended multiple power law function, defined as

$$Function\ 2)\quad \begin{cases} ax^{-b} + cx^{-d} + f & (for\ pre\text{-}edge\ region) \\ ax^{-b} + cx^{-d} + ex + f & (for\ post\text{-}edge\ region), \end{cases}$$

where *x* is the energy loss, and *a, b, c, d, e, f* are fitting parameters.



In this formulation, the linear term $ex$ in the post-edge window accounts for the contribution from the Hartree-Slater continuum. Each background function was subtracted from the raw spectrum to obtain two background-subtracted spectra.

c) To correct for plural scattering effects due to sample thickness, Fourier-ratio deconvolution was applied to each background-subtracted spectrum using the corresponding low-loss spectrum. A low-pass filter (cutoff frequencies of 0.36 eV$^{-1}$ for Mn and 0.32 eV$^{-1}$ for Co) was then applied to suppress high-frequency noise in the deconvolved spectra.

d) To enable comparison between the post-edge slope of each background-subtracted spectrum and that of the theoretical Hartree–Slater continuum model (calculated using the SIGMAL3 program)[67], a normalization procedure was applied. In this process, each spectrum was normalized so that the integrated intensity over its post-edge region matched the integral of the theoretical model within the same energy range. Linear fitting was then applied to determine the slopes of the post-edge region ($s_{post}$) for both the experimental spectrum and the theoretical Hartree–Slater continuum model ($s_{Hartree-Slater}$) within the same energy range. The slope of the pre-edge region ($s_{pre}$) was also calculated in this step.

e) Steps a)–d) were repeated across a range of fitting parameter combinations, including variations in the sizes and positions of the pre-edge and post-edge windows and the fitting weight applied during background fitting. For each combination, the resulting background-subtracted spectrum was evaluated using the following error function:

$$|s_{pre}| + C_w \times |s_{post} - s_{Hartree-Slater}|,$$

where the weight factor $C_w$ (20 for Mn and 1 for Co) was empirically chosen to minimize slope outliers and ensure consistency with the theoretical continuum slope.

f) The final background-subtracted spectrum was chosen as the one, regardless of whether it used background fitting function 1 or 2, that yielded the minimum error function and had a pre-edge slope within threshold $T_{slope}$ (0.15 eV$^{-1}$ for Mn, 0.025 eV$^{-1}$ for Co). Spectra with $|s_{pre}| > T_{slope}$ were considered unphysical and excluded.

3) EELS alignment and NMF

After drift correction and background subtraction, the same alignment parameters and binary masks obtained from the ADF tilt series were applied to align the tilt series of each energy channel in the core-loss EELS dataset. However, due to the non-uniform sample geometry, the number of electrons transmitted through the specimen varied with tilt angle, violating the tomographic summation rule. To compensate for this variation, normalization factors were applied, following a calibration procedure similar to that used for EDS normalization.

To extract spatially resolved information on the valence states of transition metals, a modified NMF method[31] was applied to the aligned and normalized EELS tilt series. The 3D EELS dataset (64 × 64 × 2048) was reshaped into a 2D matrix (4096 × 2048) and then decomposed into two non-negative matrices: **W**, representing the spatial distribution of each valence component, and **H**, containing fixed reference spectra for each valence state. Unlike conventional NMF, where both **W** and **H** are iteratively updated, our modified approach used a multiplicative update algorithm in which the spectral matrix **H** was fixed throughout the optimization.

The **H** matrices (**H**$_{Mn}$ and **H**$_{Co}$) were constructed for Mn and Co using the measured reference spectra, respectively. For Mn, **H**$_{Mn}$ contained three reference spectra corresponding to $Mn^{4+}$, $Mn^{3+}$, and $Mn^{2+}$, while for Co, **H**$_{Co}$ contained two reference spectra corresponding to $Co^{3+}$ and $Co^{2+}$. Notably, NMF is particularly well-suited for EELS analysis because EELS signals are inherently non-negative, and the physically meaningful spectral components must also satisfy this non-negativity constraint.

4) EELS reconstruction and valence state percentage estimation.

After applying the modified NMF method, the spatial component matrix (**W**) of each element was reshaped into 2D valence state maps corresponding to each tilt angle, forming an aligned tilt series for each valence state. These tilt series were then independently reconstructed into 3D tomograms using the GENFIRE algorithm with the following parameters: oversampling ratio of 2, number of iterations of 200, interpolation radius of 0.1 pixels, and discrete Fourier transform interpolation as the interpolation method. These reconstructed tomograms represent the spatial distribution of the intensity associated with each valence state but not yet properly normalized to directly provide their local fraction quantitatively.

Theoretical relationship between EELS white-line intensity (in our case, $L_2$ and $L_3$ peak) and the information about the number of unoccupied 3d orbitals for each valence state allows the proper normalization, if the contributions from continuum transitions can be properly subtracted from the spectra[64,65]. The contribution from the continuum transitions were determined by a linear background fitting to the post-edge region of each reference spectrum and then subtracted from the spectra. White-line intensities of the spectra were then integrated over the energy ranges of either 635.00 eV to 666.00 eV (for Mn) or 770.00 eV to 805.00 eV (for Co). To correct for variations in sample thickness and experimental geometry, the integrated white-line intensities were further normalized by dividing by the integrated continuum intensity, calculated as the integral of the post-edge signal over the energy range of either 705.00 eV to 755.00 eV (for Mn) or 850.00 eV to 900.00 eV (for Co).

Final normalization factors were then determined by matching the ratios of the integrated white-line intensities to the theoretical ratios of unoccupied 3d orbitals for each valence state. For Mn, the ratio of unoccupied 3d orbitals for $Mn^{4+}$, $Mn^{3+}$, and $Mn^{2+}$ is 7:6:5, and for Co, the ratio between $Co^{3+}$ and $Co^{2+}$ is 4:3. The resulting normalization factors were applied to the reconstructed tomograms to generate normalized 3D valence state maps.

**Qualitative and quantitative 3D analysis**

1) 3D visualization

The 3D structure, chemical composition, and valence state maps were visualized using the ParaView software. For 3D structure maps from ADF, the color scale limits were set to the minimum and maximum values within each volume. For the 3D chemical composition maps, the upper limits of the color scale were set to 80% and 16.7% for oxygen and transition metals, respectively. The lower limits were set to 0%. For 3D valence state maps, the upper limits of the color scale were set to 100%, 50%, 60%, 100% and 100% for $Mn^{4+}$, $Mn^{3+}$, $Mn^{2+}$, $Co^{3+}$, and $Co^{2+}$, respectively, while the lower limits were set to 0%. All 3D maps (structure, chemical composition, and valence states) for all measured particles are visualized in Supplementary Movies 1–4.

2) Surface analysis and depth profiling

We applied Otsu's method[56] to generate a 3D binary mask of each particle from the ADF tomograms. The surface region was defined as the outermost voxels of this mask. Subsequently, a series of one-voxel-thick shells was constructed by iteratively moving inward from the surface, such that the entire particle was partitioned into concentric shells of increasing depth. For surface and depth profiling, the chemical composition or valence state was calculated by averaging the values of all voxels within the defined surface region or within each shell at a given depth. These averages were first obtained for each particle individually. To obtain the values and standard deviations reported in the main figures, we performed an additional averaging across all particles under the same cycling condition.

3) Element-rich region identification

Element-rich regions were identified based on atomic percentage thresholds. For transition metals, voxels with concentrations exceeding 16.7% were classified as transition-metal-rich (Mn, Co or Ni), corresponding to the stoichiometry expected for a rock-salt structure fully depleted of lithium. Oxygen-rich regions were defined as voxels with oxygen concentrations above 80%. The volume fraction of these element-rich regions was calculated for each particle, and the values were then averaged across all particles under the same cycling condition to obtain the reported means and standard deviations.

4) Inhomogeneity analysis

To quantify the depth-dependent inhomogeneity of Mn and Co, the standard deviation of the chemical composition was calculated for each depth-dependent shell (defined during the depth-profiling as described above), yielding an inhomogeneity profile as a function of depth for each particle. Then, the calculated inhomogeneity values were averaged across all particles under the same cycling condition to obtain the reported means and standard deviations.

## Data availability

All of our experimental data, tomographic reconstructions, determined 3D structure, chemical composition and valence state maps, and electrochemical performance results will be posted on a public website upon publication.

## Code availability

Source codes will be posted on a public website upon publication. The code for multi-modal data fusion is available at https://github.com/jtschwar/tomo_TV . The software used for tilt-series alignment during the fused multi-modal tomography reconstruction is available at https://github.com/jtschwar/projection_refinement .

## Acknowledgements

We thank Hyesung Jo, Seokjo Hong, Juhyeok Lee, and Sangjun Kang for helpful discussions. This research was supported by the National Research Foundation of Korea (NRF) Grants funded by the Korean Government (MSIT) (No. RS-2023-00208179 and RS-2024–00435493). Y.Y. also acknowledges the support from the KAIST singularity professor program. R.H. acknowledges support from the U.S. Department of Energy, Basic Energy Sciences, under award DE-SC0024147. J.M. acknowledges support from the National Science Foundation GRFP. Part of the STEM experiments were conducted using a double Cs corrected Titan cubed G2 60-300 (FEI) and a Spectra Ultra TEM (Thermo Fisher) equipment at KARA. Excellent support by Hyung Bin Bae, Jin-Seok Choi, Su Min Lee, and the staff of KARA is gratefully acknowledged. The tomography data analyses were partially supported by the KAIST Quantum Research Core Facility Center (KBSI-NFEC grant funded by Korea government MSIT, PG2022004-09). We declare that the authors utilized the ChatGPT (https://chat.openai.com/chat) for language editing purpose only, and the original manuscript texts were all written by human authors, not by artificial intelligence.

## Author contributions

Y.Y. conceived the idea and directed the study. S.K. and H.B. conducted the electrochemical cycling. J.O., C.J., and Y.Y. prepared the TEM specimens. J.O., C.J., and Y.Y. designed and performed the tomography experiments. J.O., and Y.Y. conducted the experimental data analyses of electron tomography. J.M., J.S., and R.H. conducted the fused-multimodal tomography. J.O., S.K., and Y.Y. wrote the manuscript. All authors commented on the manuscript.

## Competing interests

J.O., C.J., and Y.Y. have a patent application (Korea, 10-2025-0120022), which disclose the methods for simultaneous measurement of the 3D structure, chemical composition and valence states. The remaining authors declare no competing interests.

## Additional information

Correspondence and requests for materials should be addressed to H.B. (hrbyon@kaist.ac.kr), and Y.Y. (yongsoo.yang@kaist.ac.kr)

# Supplementary Information

for

Correlative 3D Mapping of Structure, Composition, and Valence State Dynamics in Battery Cathodes via Simultaneous ADF-EDS-EELS Tomography

Jaewhan Oh[1], Sunggu Kim[2], Chaehwa Jeong[1], Jason Manassa[3], Jonathan Schwartz[3,4], Sangmoon Yoon[5], Robert Hovden[3], Hye Ryung Byon[2†] and Yongsoo Yang[1,6*]

[1]*Department of Physics, Korea Advanced Institute of Science and Technology (KAIST), Daejeon, Republic of Korea*
[2]*Department of Chemistry, Korea Advanced Institute of Science and Technology (KAIST), Daejeon, Republic of Korea*
[3]*Department of Materials Science and Engineering, University of Michigan, Ann Arbor, MI, USA*
[4]*Chan Zuckerberg Imaging Institute, Redwood City, CA, USA*
[5]*Department of Physics and Semiconductor Science, Gachon University, Seongnam, South Korea*
[6]*Graduate School of Semiconductor Technology, School of Electrical Engineering, Korea Advanced Institute of Science and Technology (KAIST), Daejeon, Republic of Korea*

*Corresponding author, email: [†]hrbyon@kaist.ac.kr, [*]yongsoo.yang@kaist.ac.kr

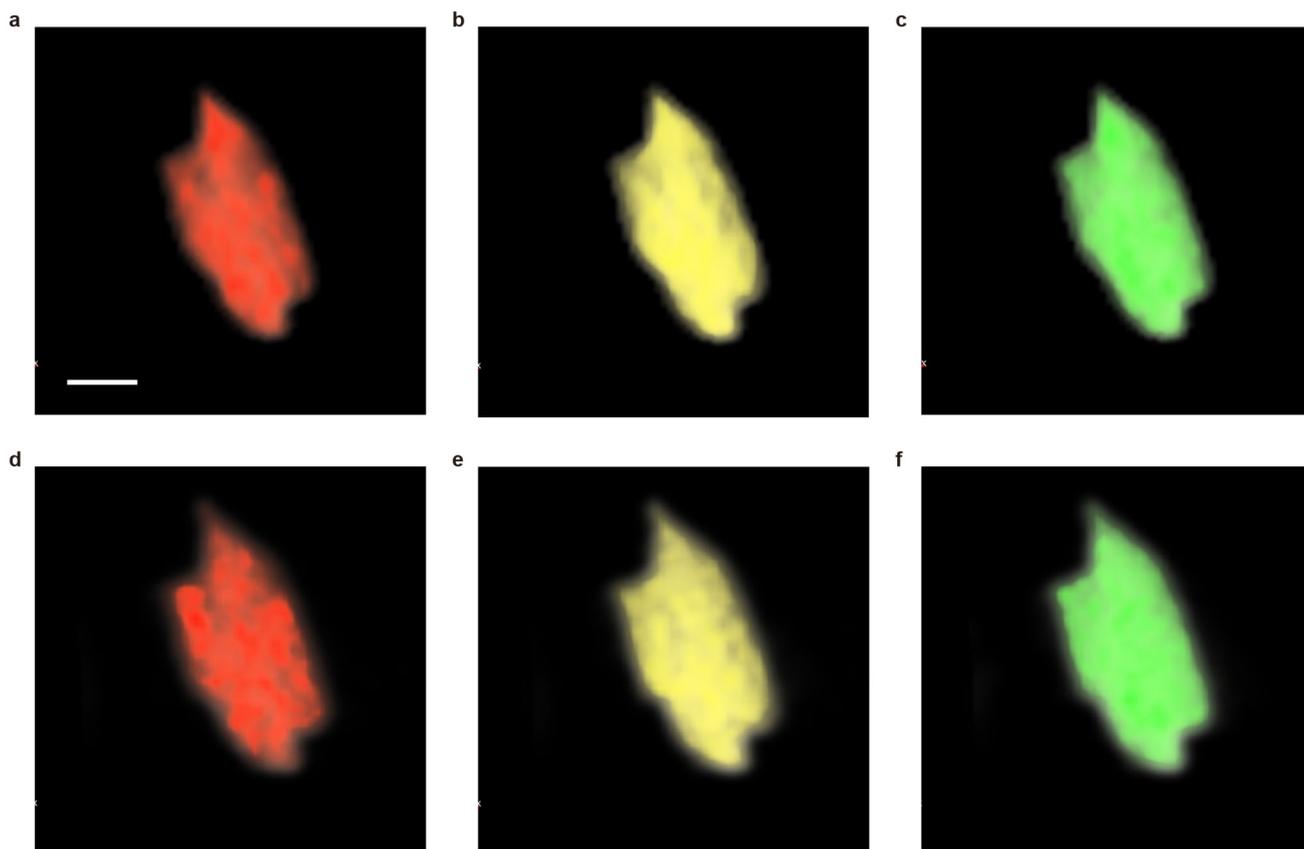

**Supplementary Figure 1 | Comparison between two different EDS reconstruction methods. a–c,** 3D rendering of EDS reconstructions of Mn **(a)**, Co **(b)**, and Ni **(c)** obtained from direct reconstruction using the simultaneous ADF-EDS-EELS tomography method. **d–f,** EDS reconstructions of Mn **(d)**, Co **(e)**, and Ni **(f)** obtained using the fused multi-modal tomography method. Scale bar, 50 nm.

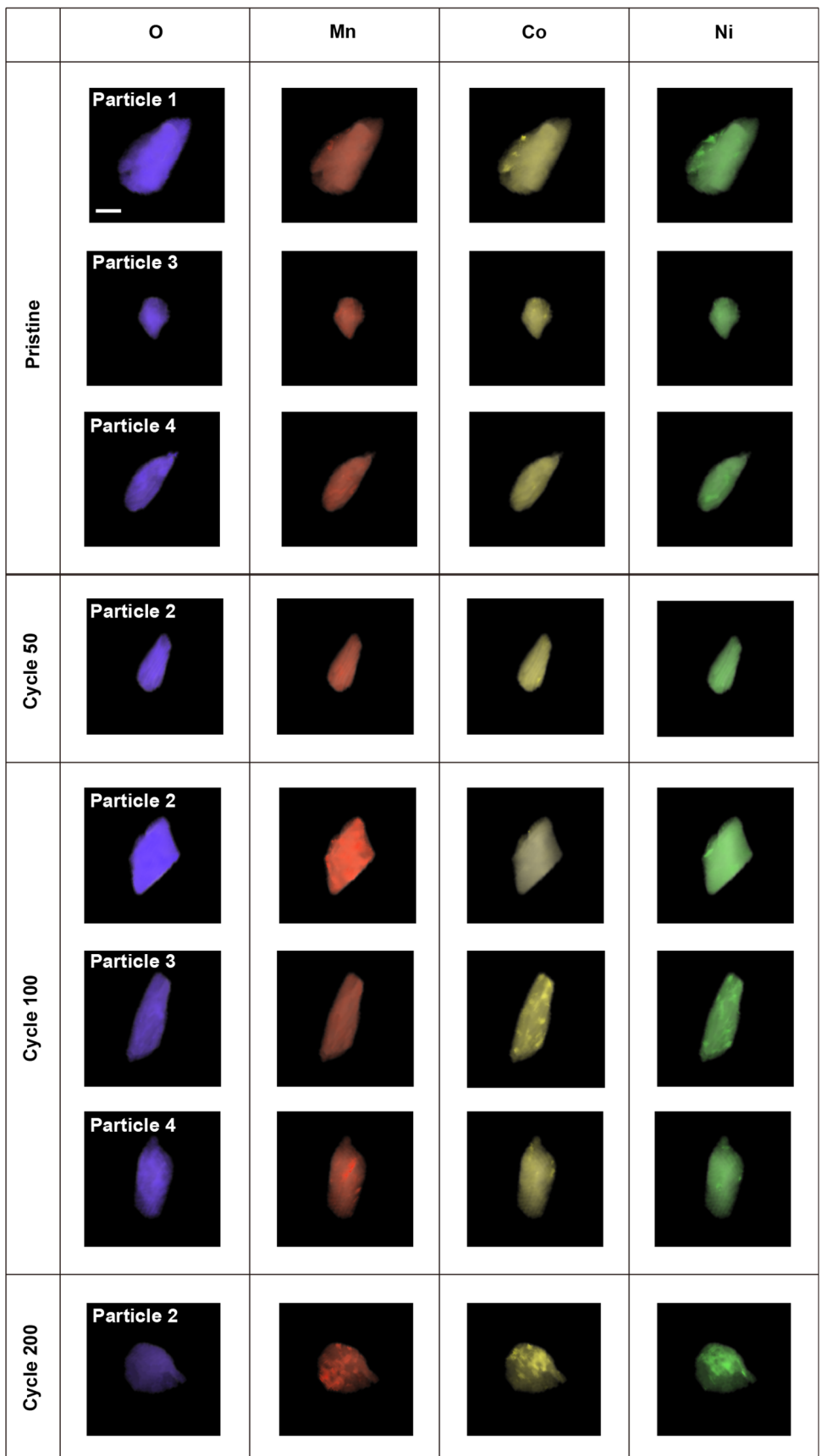

**Supplementary Figure 2 | 3D rendering of chemical composition maps of oxygen and transition metals for NCM111 particles at different electrochemical cycles.** These maps display the 3D chemical distributions of O, Mn, Co, and Ni for the particles not shown in Fig. 2 (three pristine, one 50-cycled, three 100-cycled, and one 200-cycled). Each column corresponds to a different element, and each row represents an individual particle. Scale bar, 50 nm.

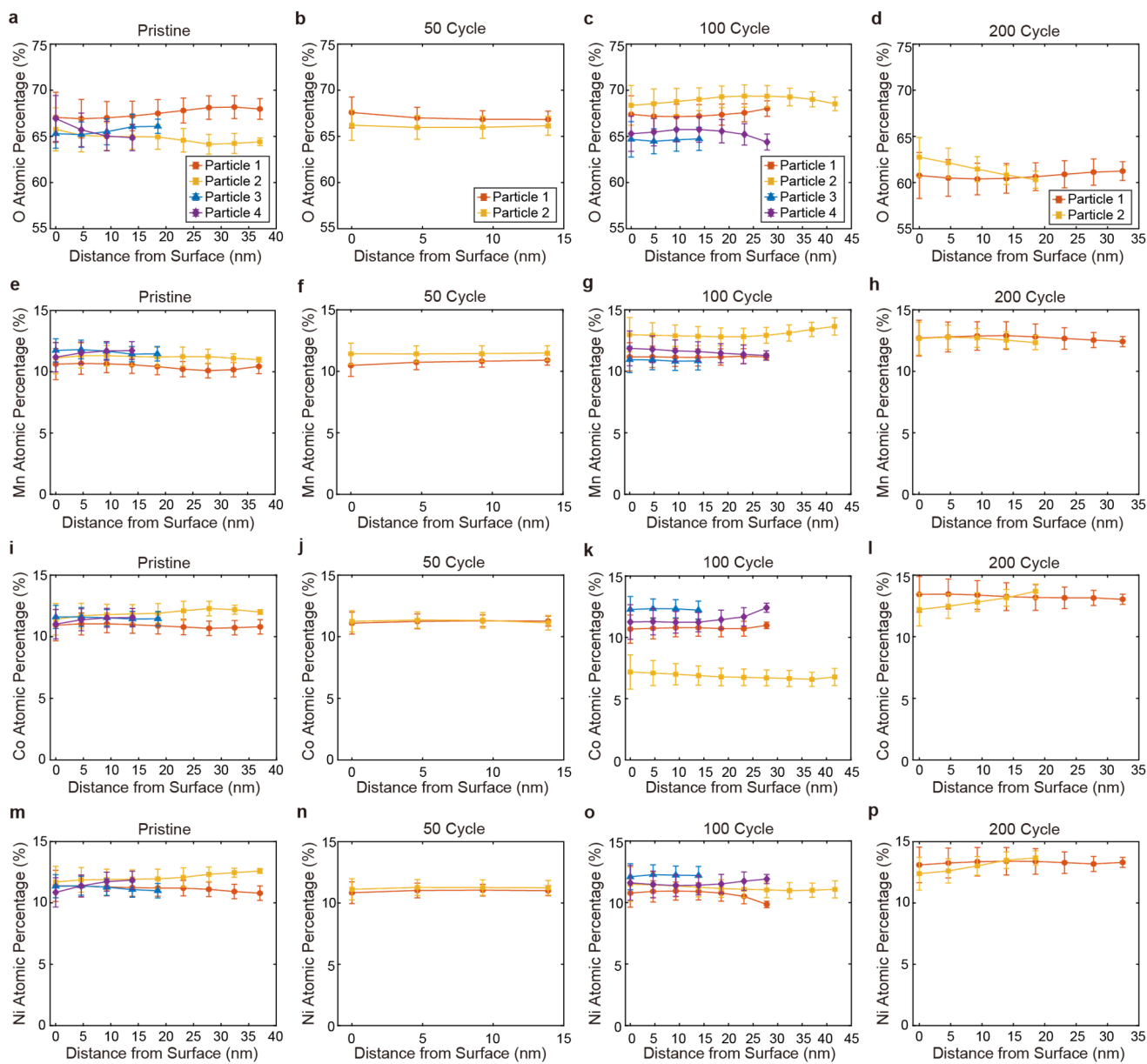

**Supplementary Figure 3 | Depth profiles of atomic composition for all NCM111 particles analyzed in this study. a–d,** Depth profiles of O chemical composition for pristine **(a)**, 50-cycled **(b)**, 100-cycled **(c)**, and 200-cycled **(d)** particles. **e–h,** Depth profiles of Mn chemical composition for pristine **(e)**, 50-cycled **(f)**, 100-cycled **(g)**, and 200-cycled **(h)** particles. **i–l,** Depth profiles of Co chemical composition for pristine **(i)**, 50-cycled **(j)**, 100-cycled **(k)**, and 200-cycled **(l)** particles. **m–p,** Depth profiles of Ni chemical composition for pristine **(m)**, 50-cycled **(n)**, 100-cycled **(o)**, and 200-cycled **(p)** particles. All error bars represent standard deviations at each depth interval, indicating variation in atomic composition at the corresponding depth.

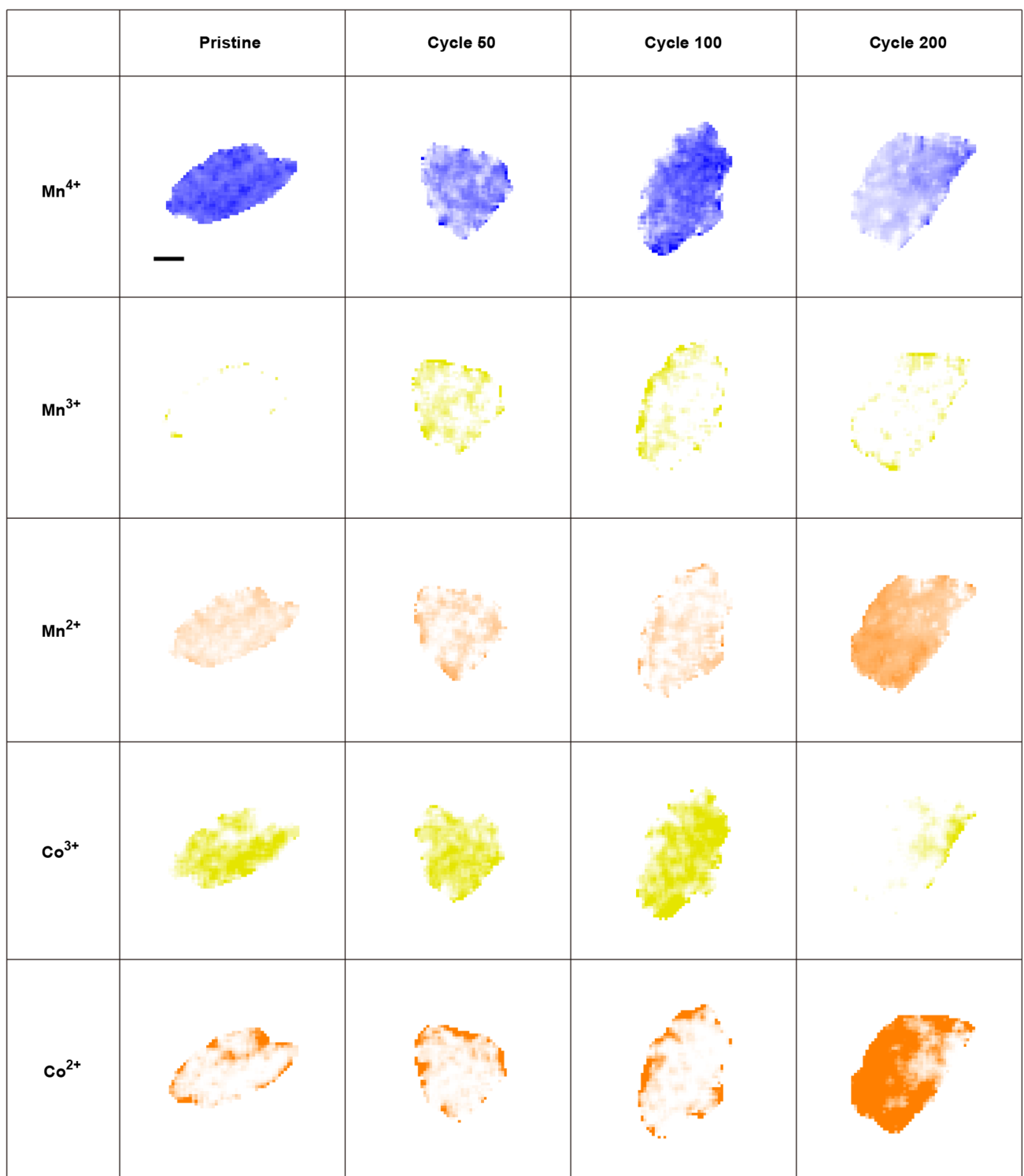

**Supplementary Figure 4 | 2D EELS analysis for evolution of Mn and Co valence states during electrochemical cycling.** 2D valence state maps of $Mn^{4+}$, $Mn^{3+}$, $Mn^{2+}$, $Co^{3+}$, and $Co^{2+}$ from representative particles at different electrochemical cycles: particle 1 from the pristine samples, particle 2 from the 50-cycled samples, particle 1 from the 100-cycled samples, and particle 1 from the 200-cycled samples. Each column corresponds to a different electrochemical cycle, and each row represents a different valence state. Clear changes in the spatial distribution of valence states were observed between the pristine and 200-cycled particles. Scale bar, 50 nm.

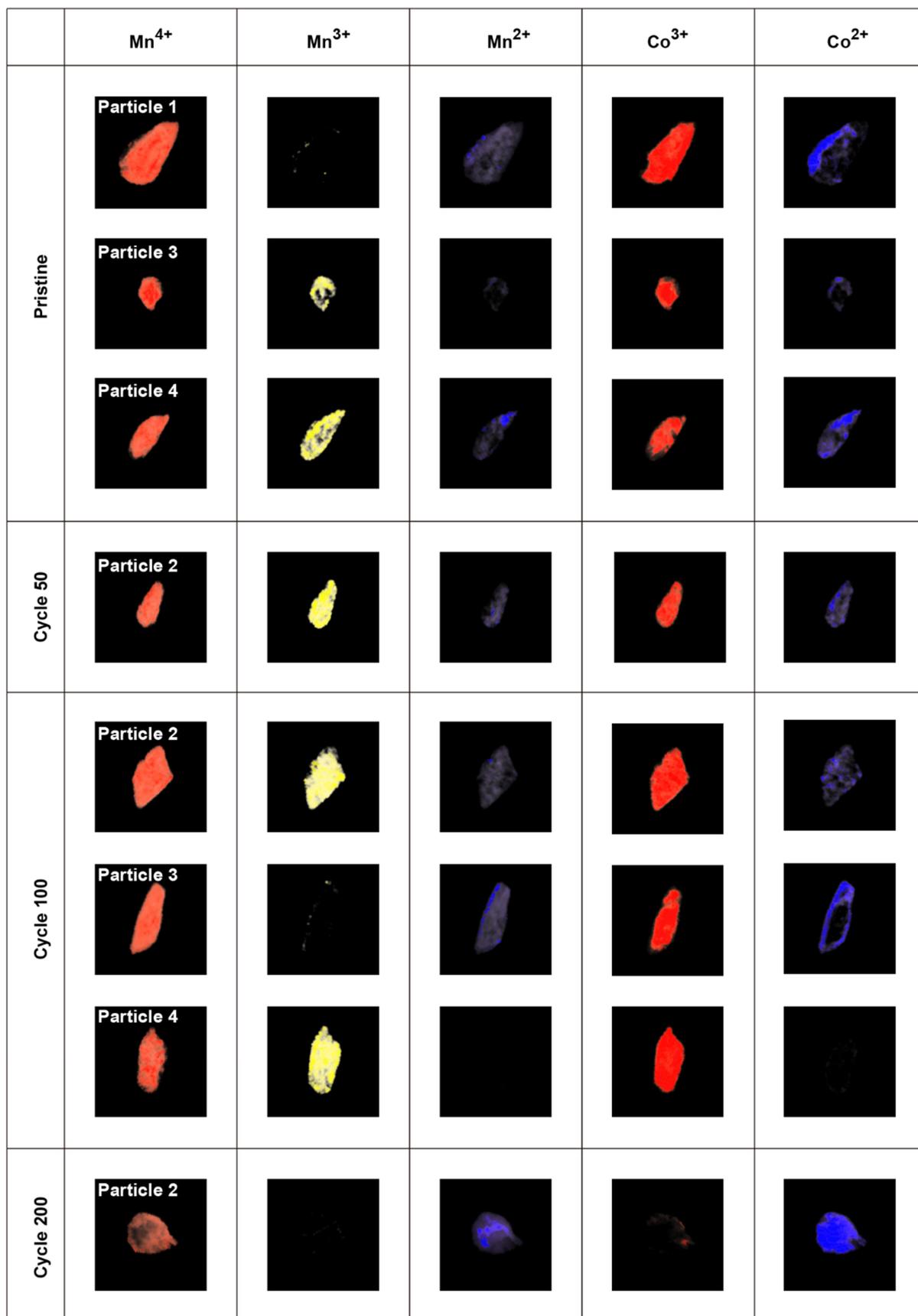

**Supplementary Figure 5 | 3D rendering of valence state maps of Mn and Co for NCM111 particles at different electrochemical cycles.** These maps show the 3D valence state distributions of $Mn^{4+}$, $Mn^{3+}$, $Mn^{2+}$, $Co^{3+}$, and $Co^{2+}$ for the particles not shown in Fig. 3 (three pristine, one 50-cycled, three 100-cycled, and one 200-cycled). Each column represents a different valence state, and each row corresponds to an individual particle. Scale bar, 50 nm.

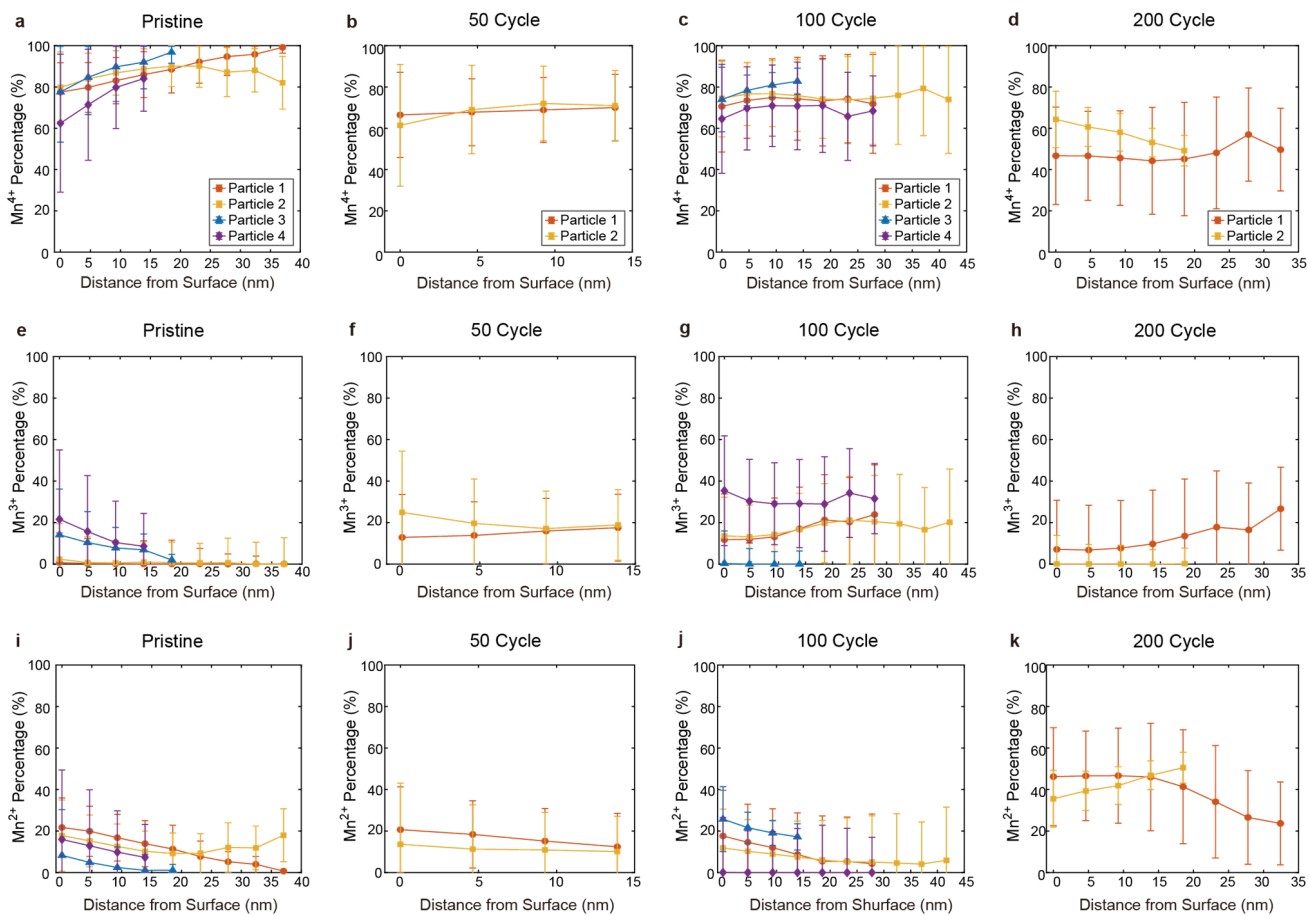

**Supplementary Figure 6 | Depth profiles of Mn valence states for all NCM111 particles analyzed in this study. a–d,** Depth profiles of the $Mn^{4+}$ for pristine **(a)**, 50-cycled **(b)**, 100-cycled **(c)**, and 200-cycled **(d)** particles. **e–h,** Depth profiles of the $Mn^{3+}$ for pristine **(e)**, 50-cycled **(f)**, 100-cycled **(g)**, and 200-cycled **(h)** particles. **i–l,** Depth profiles of the $Mn^{2+}$ for pristine **(i)**, 50-cycled **(j)**, 100-cycled **(k)**, and 200-cycled **(l)** particles. All error bars represent standard deviations at each depth interval, indicating variation in Mn valence state percentages.

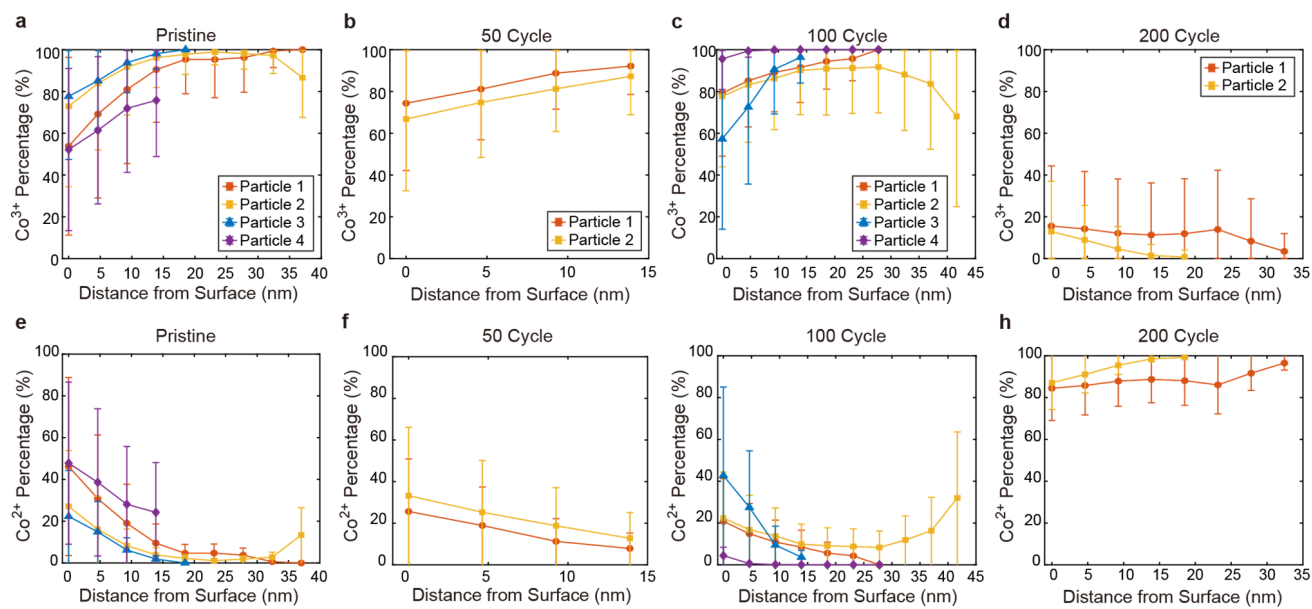

**Supplementary Figure 7 | Depth profiles of Co valence states for all NCM111 particles analyzed in this study. a–d,** Depth profiles of $Co^{3+}$ for pristine **(a)**, 50-cycled **(b)**, 100-cycled **(c)**, and 200-cycled **(d)** particles. **e–h,** Depth profiles of $Co^{2+}$ for pristine **(e)**, 50-cycled **(f)**, 100-cycled **(g)**, and 200-cycled **(h)** particles. All error bars represent standard deviations at each depth interval, indicating variation in Co valence state percentages.

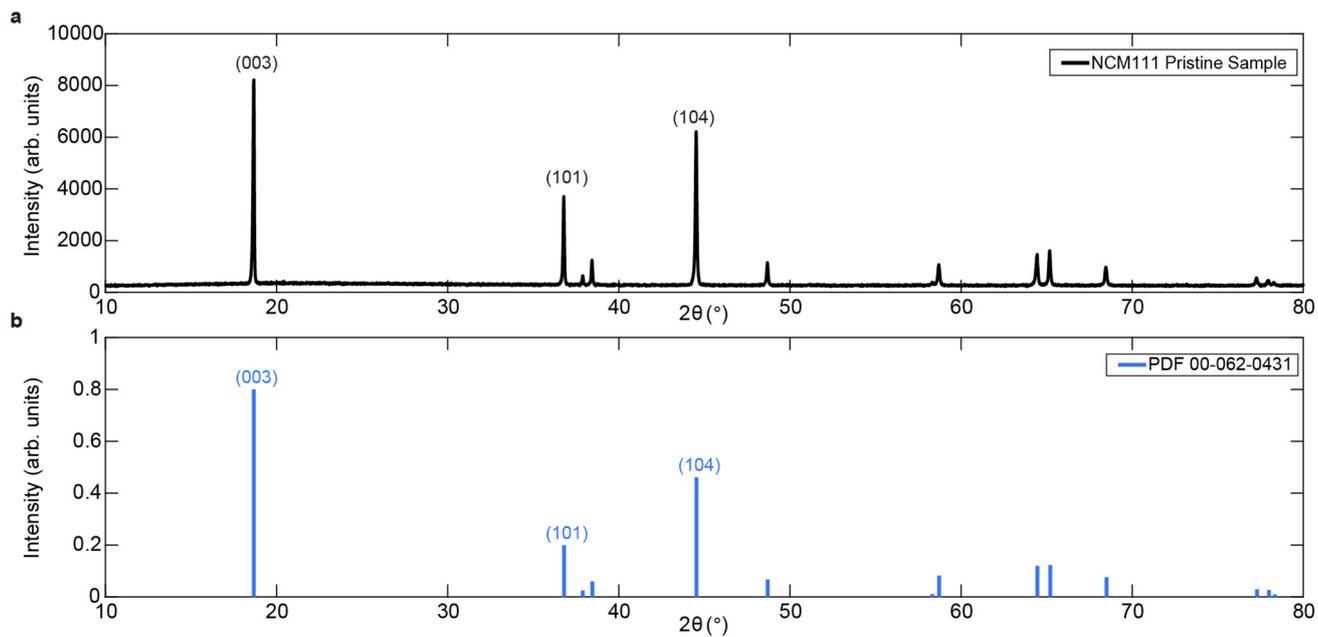

**Supplementary Figure 8 | Powder XRD analysis of pristine NCM111 sample. a,** Experimental powder XRD pattern of the pristine NCM111 sample. **b,** Reference pattern from the standard database (PDF 00-062-0431). The experimental pattern shows a strong match in peak positions with the reference, confirming that the pristine NCM111 sample exhibits a well-ordered layered structure without any detectable secondary phases.

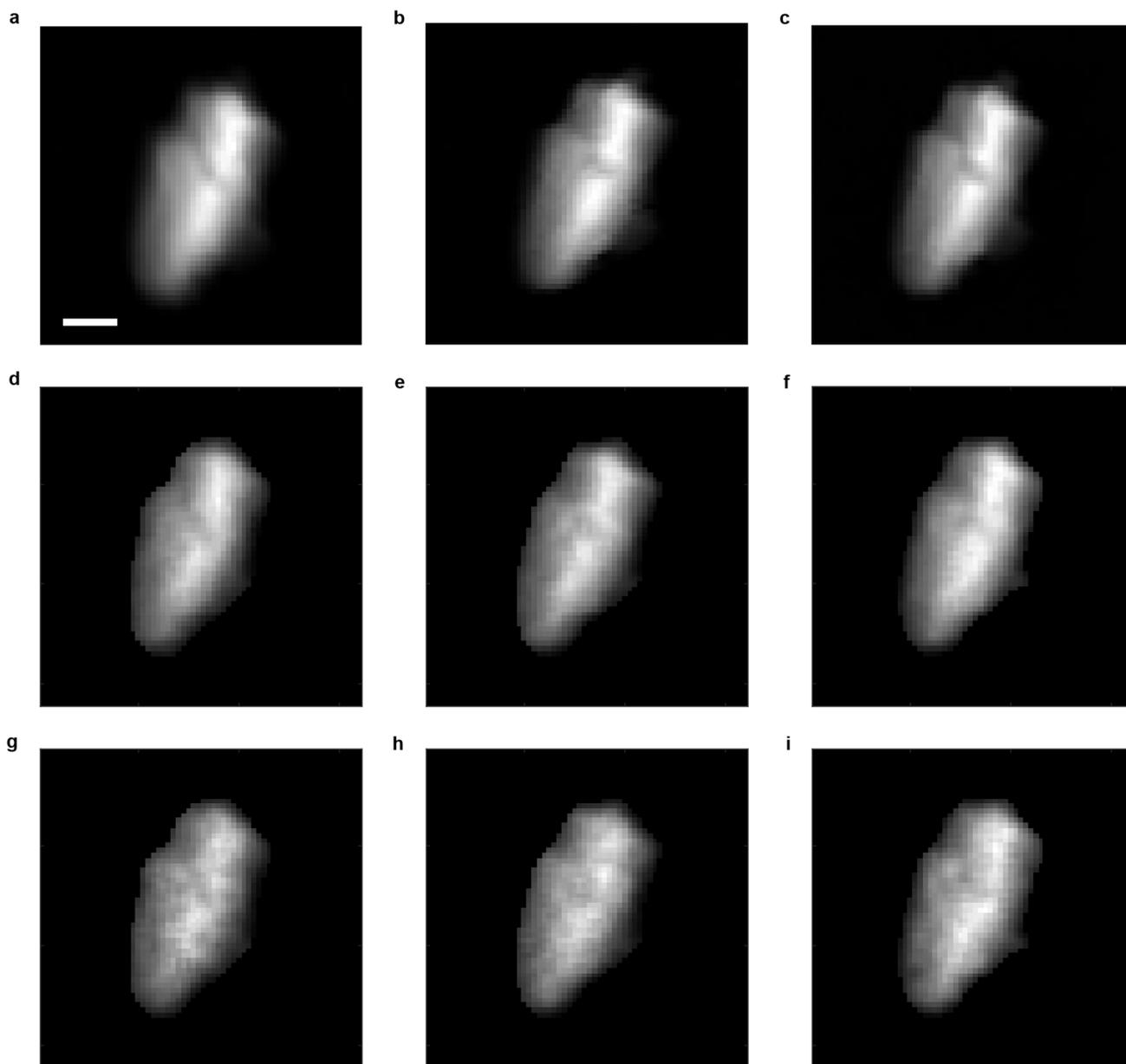

**Supplementary Figure 9 | Verification of beam stability: ADF and EELS images before, in the middle, and after tomography acquisition, showing no discernible structural change. a**–**c**, Zero-degree ADF images taken at the beginning **(a)**, in the middle **(b)**, and at the end **(c)** of the tilt series acquisition. **d**–**f,** Zero-degree EELS Mn $L_3$ edge intensity maps taken at the beginning **(d)**, in the middle **(e)**, and at the end **(f)** of the tilt series acquisition. **g**–**i,** Zero-degree EELS Co $L_3$ edge intensity maps taken at the beginning **(g)**, in the middle **(h)**, and at the end **(i)** of the tilt series acquisition. Scale bar, 50 nm.

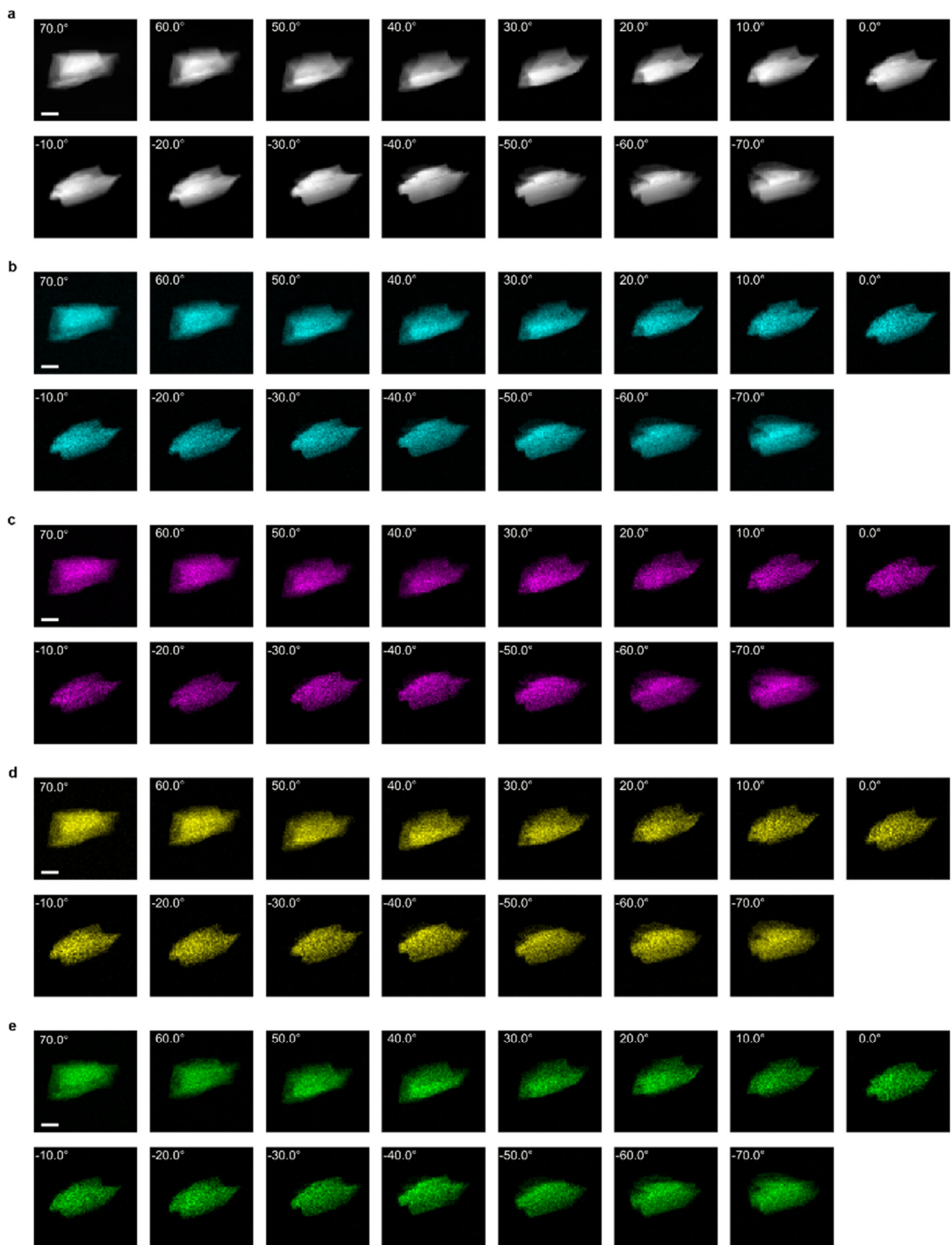

**Supplementary Figure 10 | A representative tomographic tilt series images for particle 1 of the pristine NCM111 sample. a,** ADF tilt series. **b–e,** EDS tilt series of O **(b)**, Mn **(c)**, Co **(d)**, and Ni **(e)**. Scale bar, 50 nm.

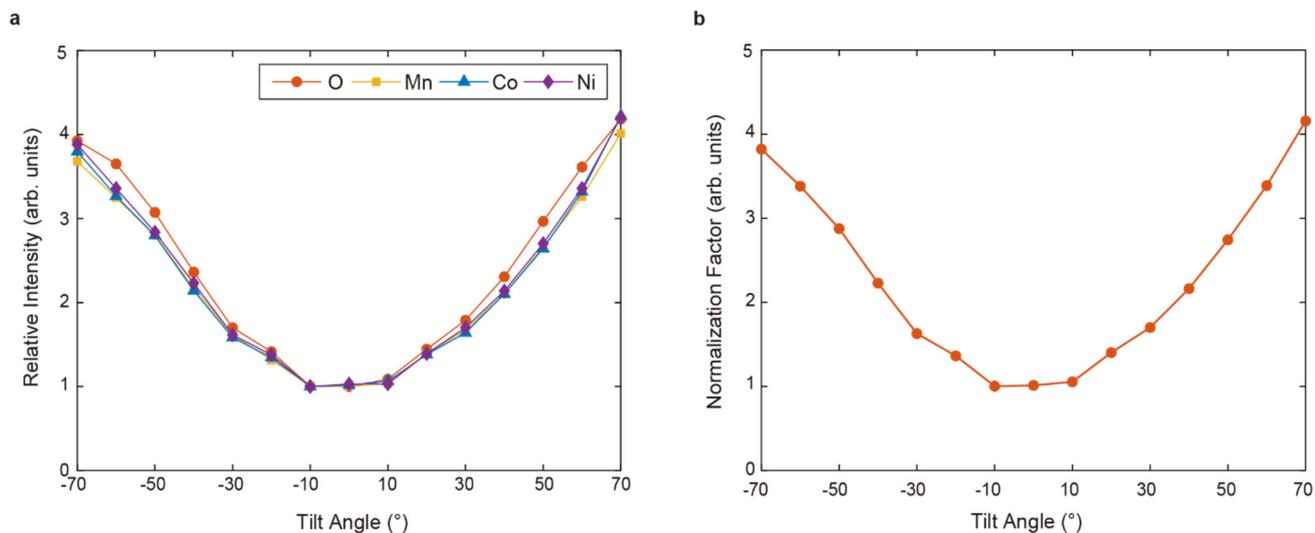

**Supplementary Figure 11 | Calibration of angle-dependent X-ray absorption effects for EDS acquisition. a,** Relative intensity of oxygen and transition metal EDS signals as a function of tilt angle. **b,** Averaged normalization factors determined from the data shown in **(a)**.

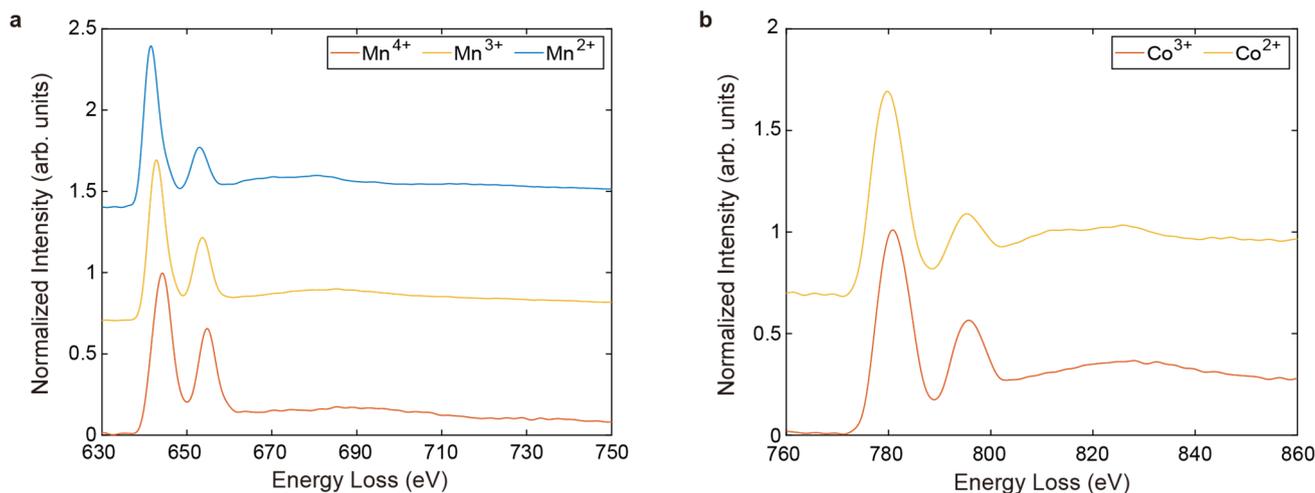

**Supplementary Figure 12 | Reference EELS spectra of Mn and Co valence states. a,** Reference spectra for $Mn^{4+}$, $Mn^{3+}$, and $Mn^{2+}$ obtained from $MnO_2$, $Mn_2O_3$, and MnO samples, respectively. **b,** Reference spectra for $Co^{3+}$ and $Co^{2+}$ obtained from $LiCoO_2$ and CoO samples, respectively.

|  | Li | Mn | Co | Ni |
|---|---|---|---|---|
| Pristine | 49.3% | 16.9% | 16.8% | 17.0% |

**Supplementary Table 1 | ICP-OES analysis of atomic composition in pristine NCM111 sample.** Atomic percentages of Li, Mn, Co, and Ni in the pristine NCM111 sample were quantified using ICP-OES. The results confirm the expected stoichiometry of pristine NCM111.

**Supplementary Movie 1 | 3D rendering of chemical composition maps of oxygen and transition metals for NCM111 particles at different electrochemical cycles**. These maps display the 3D chemical distributions of O, Mn, Co, and Ni for all particles used in this study (four pristine, two 50-cycled, four 100-cycled, and two 200-cycled). Each column corresponds to a different electrochemical cycle, and each row represents an individual particle. Scale bar, 50 nm.

**Supplementary Movie 2 | 3D rendering of valence state maps of Mn for NCM111 particles at different electrochemical cycles**. These maps display the 3D distributions of Mn4+, Mn3+, and Mn2+ for all particles used in this study (four pristine, two 50-cycled, four 100-cycled, and two 200-cycled). Each column corresponds to a different electrochemical cycle, and each row represents an individual particle. Scale bar, 50 nm.

**Supplementary Movie 3 | 3D rendering of valence state maps of Co for NCM111 particles at different electrochemical cycles**. These maps display the 3D distributions of Co3+, and Co2+ for all particles used in this study (four pristine, two 50-cycled, four 100-cycled, and two 200-cycled). Each column corresponds to a different electrochemical cycle, and each row represents an individual particle. Scale bar, 50 nm.

**Supplementary Movie 4 | 3D rendering of structure maps from ADF tomography for NCM111 particles at different electrochemical cycles**. These maps display the 3D structural features of all particles used in this study (four pristine, two 50-cycled, four 100-cycled, and two 200-cycled). Each column corresponds to a different electrochemical cycle, and each row represents an individual particle. Scale bar, 50 nm.